\documentclass{article}
\usepackage{arxiv}
\usepackage{booktabs}
\usepackage{tabularx}
\usepackage{enumitem}
\usepackage{tabularx}
\usepackage{booktabs}
\usepackage{ragged2e}
\usepackage{array}               
\usepackage{tabularx}
\usepackage{url}
\usepackage[sorting=none]{biblatex}
\usepackage{enumitem}
  
\renewcommand{\arraystretch}{1.2}

\usepackage{graphicx}%
\usepackage{multirow}%
\usepackage{amsmath,amssymb,amsfonts}%
\usepackage{amsthm}%
\usepackage{mathrsfs}%
\usepackage{xcolor}%
\usepackage{braket}
\usepackage{textcomp}%
\usepackage{manyfoot}%
\usepackage{booktabs}%
\usepackage{algorithm}%
\usepackage{algorithmicx}%
\usepackage{algpseudocode}%
\usepackage{listings}%
\usepackage{tcolorbox}
\usepackage{float}
\usepackage[utf8]{inputenc} 
\usepackage[T1]{fontenc}    
\usepackage{hyperref}       
\usepackage{url}            
\usepackage{booktabs}       
\usepackage{amsfonts}       
\usepackage{nicefrac}       
\usepackage{microtype}      
\usepackage{lipsum}
\usepackage{graphicx}
\graphicspath{ {./images/} }
\title{Quantum Key Distribution: Bridging Theoretical Security Proofs, Practical Attacks, and Error Correction for Quantum-Augmented Networks}

\author{
 Nitin Jha \\
  Department of Computer Science\\
  Kennesaw State University\\
  Marietta, GA, 30066 \\
  \texttt{njha1@students.kennesaw.edu}\\
   \And
 Abhishek Parakh \\
  Department of Computer Science\\
  Kennesaw State University\\
  Marietta, GA, 30066 \\
  \texttt{aparakh@kennesaw.edu} \\
  \And
 Mahadevan Subramaniam \\
  Department of Computer Science\\
  University of Nebraska Omaha\\
  Omaha, NE 68182 \\
  \texttt{msubramaniam@unomaha.edu} \\
}

\begin{document}
\maketitle
\begin{abstract}
Quantum Key Distribution (QKD) is revolutionizing cryptography by promising information-theoretic security through the immutable laws of quantum mechanics. Yet, the challenge of transforming these idealized security models into practical, resilient systems remains a pressing issue, especially as quantum computing evolves. In this review, we critically dissect and synthesize the latest advancements in QKD protocols and their security vulnerabilities, with a strong emphasis on rigorous security proofs. We actively categorize contemporary QKD schemes into three key classes: uncertainty principle-based protocols (e.g., BB84), hybrid architectures that enable secure direct communication (eg, three-stage protocol), and continuous-variable frameworks. We further include two modern classes of QKD protocols, namely Twin-field QKD and Device-Independent QKD, both of which were developed to have practical implementations over the last decade. Moreover, we highlight important experimental breakthroughs and innovative mitigation strategies, including the deployment of advanced Quantum Error Correction Codes (QECCs), that significantly enhance channel fidelity and system robustness. {By mapping the current landscape, from sophisticated quantum attacks to state-of-the-art error correction methods, this review fills an important gap in the literature. To bring everything together, the relevance of this review concerning quantum augmented networks (QuANets) is also presented. This allows the readers to gain a comprehensive understanding of the security promises of quantum key distribution from theoretical proofs to experimental validations}.
\end{abstract}


\section{Introduction}
From Caesar’s shift cipher to modern-day encryption algorithms, the need for private communication has driven cryptographic innovation for centuries. In traditional cryptography methods, two legitimate parties, Alice and Bob, share a communication channel and communicate privately even in the presence of an illegitimate party, Eve.  Our current standards of secure communications are derived from the last seven decades of information science and cryptography schemes such as public-key authentication. However, in the last three decades, the introduction of quantum computers has put the security of current communications in need of serious revision. 

Quantum Key Distribution (QKD) is one of the major candidates for secure communication in the age of quantum computers. QKD establishes shared secret keys between two parties, Alice and Bob, by exploiting the principles of quantum mechanics  \cite{Gisin2002, Scarani2009}.   Throughout the last couple of decades, the development of QKD networks has made great strides. Intercity networks such as the DARPA network  \cite{darpanetwork}, Trieste network  \cite{trieste2023}, Tokyo network  \cite{tokyonetwork}, etc., are prime examples of rapidly growing interest in the development of practical quantum networks. In addition to land-based quantum networks, the Chinese Academy of Sciences (CAS) developed a satellite-based quantum network using a decoy-state QKD transmitter on a low-orbit satellite  \cite{cas}.  The rapidly growing interest in QKD calls for the evaluation of its security measures.

The security of QKD is due to the inherent laws of physics  \cite{bennett1984update, ekert1991quantum}. The laws, such as the \textit{no-cloning} principle, which dictates that unknown quantum states cannot be cloned with absolute certainty  \cite{nocloning}.  This establishes the unconditional security basis for QKD. Therefore, the use of an ideal single-photon source to transmit a quantum state to generate a key eliminates any chances that the eavesdropper could clone the state and obtain the key for themselves, thus establishing a theoretically secure key. However, single-photon sources are not the most practical in the current times and have many associated problems, such as random bursts of multiple photons \cite{singlephoton}. There are several other issues related to the security and overall efficiency of QKD protocols, one such problem being noisy communication channels. In general, we cannot achieve 100\% security for any key distribution system whose security depends on physical devices. This is due to several imperfections in physical devices and their implementations  \cite{security}. The effect of channel noise cannot be distinguished from an eavesdropper attack, and thus, it presents a serious problem. Even with quantum error correction codes (QECC), we still need to estimate the possible information leaked to the eavesdropper. This requires modification of existing security proofs of QKD protocols to account for several channel noises.

{In this review, we survey quantum key distribution (QKD) protocols from legacy schemes (e.g., BB84, B92) to modern designs (e.g., twin-field), covering both theoretical foundations and implementation details. The article is intended for readers with basic familiarity with quantum networking and maps protocol choices to their associated threat models by exploring practical attacks that can compromise security. We also assess QKD’s security guarantees through standard mathematical proof frameworks and, finally, review quantum error-correction codes (QECCs), at both the theoretical and implementation point of view. We also include a discussion about how these attack and defense mechanics are essential to efficient working of quantum-augmented networks (QuANets), which can serve as the long-term solution to deploy large scale quantum-augmented networks.}

This work is organized as follows: Sec \ref{Sec:Scope} explains the scope of this review article. Sec \ref{QKD} explains the basics of different QKD protocols. This section is divided into three subsections: protocols based on the uncertainty principle (BB84 protocol), protocols that can be used for secure communication (three-stage protocol), and continuous variable protocols. Sec \ref{Attacks}  explores several practical attacks on the QKD system, and highlights some of the experimental and simulation-based works on the Photon-number splitting attack, the Trojan horse attack, and jamming attacks. Sec \ref{Security-analysis}  goes over the security paradigms in QKD systems, summarizes some of the standard security proofs for the several QKD classes mentioned above. Sec \ref{QECC}  discusses some of the benchmark quantum error correction codes, along with some recent theoretical and experimental works realizing several QECCs. Sec \ref{Sec:Relevance} discusses the relevance of this review work in the current age of quantum networks and how this analysis can be of importance in the discussion of quantum-augmented networks (QuANets). 

\subsection{Scope of this Review}
\label{Sec:Scope}

The field of \textit{quantum communication} has seen massive strides in recent years. Several practical networks have been constructed, and several practical attacks have been explored using both simulators and experimental setups. There have been several review works done on QKD and quantum cryptography. Xu et al.,  \cite{xu2020secure} present an in-depth review of QKD strategies and different hacking attempts on QKD networks. The work explores different theoretical security proofs associated with QKD protocols, such as Shor-Preskill  \cite{shor1995scheme}, and Lo-Chau  \cite{lo1999unconditional} security proofs. However, this work does not review the security proofs for the continuous-variable QKD protocol. Apart from this, the review work was done in $2020$, and there have been significant developments regarding different attacks on quantum networks. In another study, Horace et al., \cite{security} explored the core information-theoretic security guarantees of cryptosystems and revealed significant limitations in existing QKD security analyses. This work, too, does not explore experimental updates regarding the security of QKD protocols. The main focus of this review is to fill this gap in some significant review works regarding the security of QKD protocols, experimental works on attacks on QKD protocols, and error correction codes. 

The focus of this review lies in both the theoretical and practical implementations, which are extremely necessary for the development of global-scale quantum networks. 

\begin{enumerate}
    \item There has been groundbreaking theoretical work on the security of quantum networks. However, most security proofs were developed under idealized assumptions during the early stages of quantum‐computing research. Now that experimental capabilities have advanced, these foundational proofs must be revisited and re‐examined in light of the many new attack models proposed in recent years.
    \item From an experimental and simulation standpoint, several new and more accessible platforms now allow quantum protocols to be tested against standard attacks. These small-scale experimental results—i.e., performance under various attack models—are crucial for discussions of quantum-internet deployment. Moreover, reviewing theoretical and experimental foundations in parallel is essential for designing new protocols and helps bridge the gap between idealized theory and practical implementation.

    \item The discussion of quantum error‐correction codes, though often presented as abstract, theoretical constructs, directly addresses the noise and decoherence plaguing today’s quantum hardware. Yet most surveys of the security of quantum networks sidestep QECCs, even though they are vital for preserving the integrity of transmitted quantum states. By detecting and correcting both bit-flip and phase-flip errors, QECCs ensure that the low error rates assumed in security proofs truly reflect adversarial tampering rather than device imperfections. Moreover, in repeater‐based architectures and device-independent QKD, logical encoding via QECCs is the only practical path to long‐distance, high-fidelity entanglement and the low logical error thresholds these protocols demand.
\end{enumerate}
Throughout this review, we will explore some of the experiments conducted in recent times that focus on several attacks that compromise both the security and quality of transmission. We also review fundamental formulation and recent works on QECCs, as these are integral to the discussion of secure transmission in a noisy quantum channel, as also discussed in \cite{jha2024joint}.

\section{Quantum Key Distribution}
\label{QKD}
Quantum Key Distribution (QKD) utilizes quantum mechanics to provide information-theoretic security  \cite{bennett1984update, ekert1991quantum}. Alice and Bob exchange quantum states over an unauthenticated channel and then use an authenticated classical channel to reconcile their measurements. Any attempt by Eve to intercept these states inevitably disturbs them, raising the observed error rate; if it exceeds a threshold, the key is discarded and the protocol restarts. Because QKD’s security arises from physics rather than computational assumptions, an eavesdropper must attack in real time; retrospective cryptanalysis is impossible.

\begin{figure*}[ht!]
    \centering
    \includegraphics[width=0.7\linewidth]{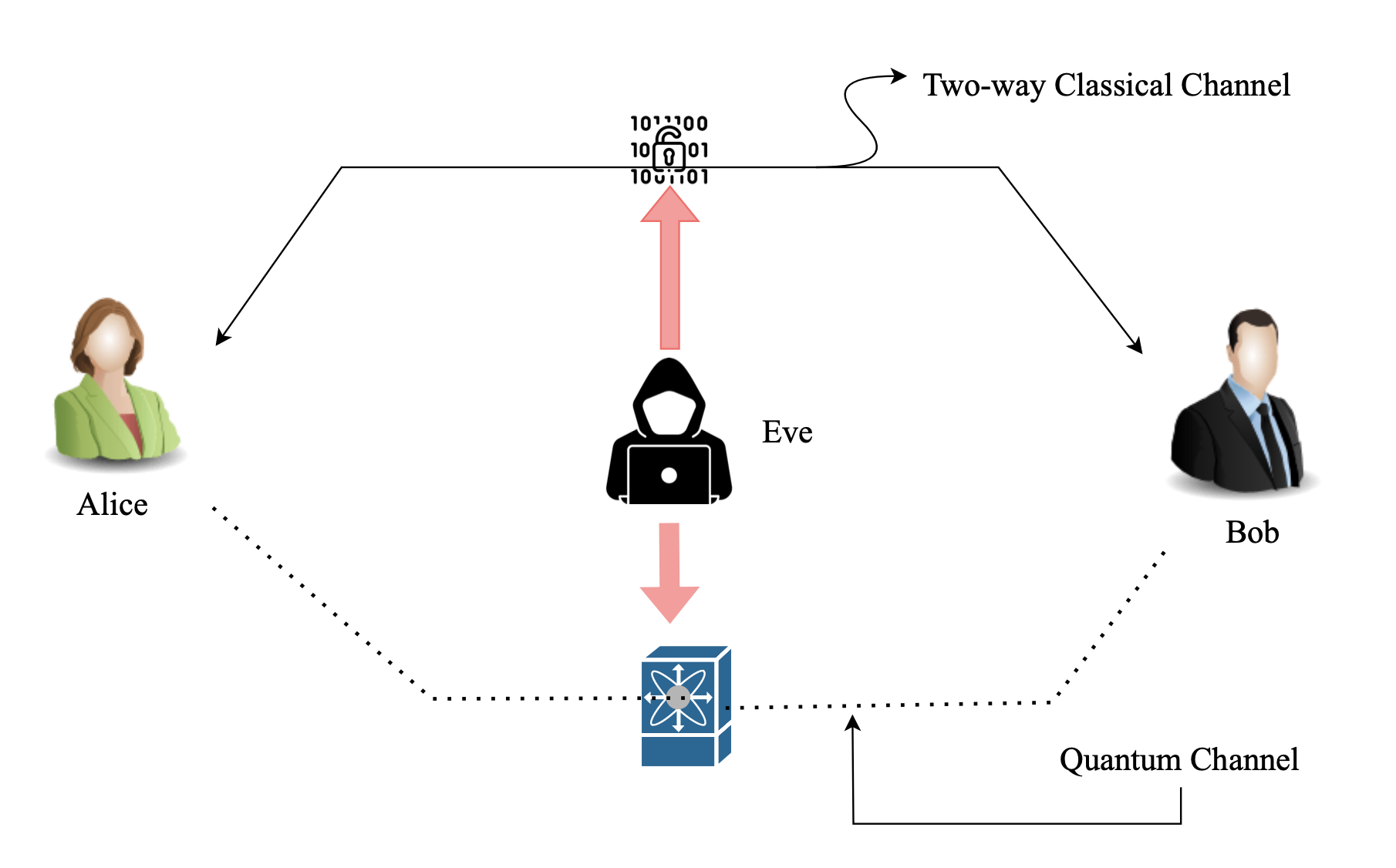}
    \caption{A schematic representation of a general QKD scenario. Alice is the sender, Bob is the receiver, and Eve is the eavesdropper present in the setup. A two-way classical communication channel and a quantum channel connect Alice and Bob. Eve can perform an eavesdropper attack on either or both of the channels to gain maximum knowledge without getting exposed. }
    \label{fig:qkd}
\end{figure*}

The no-cloning theorem ensures (1) any eavesdropper introduces detectable disturbances, and (2) Alice and Bob can bound Eve’s information to determine the amount of privacy amplification needed  \cite{yang2015trojan}. While this gives unconditional security in theory, real devices suffer side channels and implementation flaws that Eve can exploit.

QKD protocols fall into two broad categories,
\begin{enumerate}
    \item In single-photon QKD protocols—most notably BB84—each pulse ideally contains one photon whose polarization encodes a bit \cite{chan2015multiphoton}.  Practical systems use weak coherent pulses with decoy-state methods to approximate true single photons, which reduces source complexity but also lowers key rates.  Early experiments achieved $1.9$ Mb/s over $10$ km ($2$ dB loss) \cite{dynes2016ultra} and recent work reports $13.72$ Mb/s at 2 dB loss \cite{yuan201810} and up to 107 Mb/s over a 37-core fiber ($65$ Mb/s alongside $370$ Gb/s of classical data) \cite{8527341}.  By comparison, classical key-exchange can exceed $0.75$ Gb/s \cite{gao20210}, highlighting that discrete-variable QKD remains orders of magnitude slower and still susceptible to photon-number-splitting attacks \cite{harun2018evaluation}.

    \item In a multi-photon approach, there are multiple photons emitted per laser pulse. The single-photon approach is susceptible to PNS attack, as mentioned earlier, due to irregularities in the number of photons emitted in a light pulse. In theory, multiphoton protocols should offer higher transmission distances and higher transmission rates than single-photon protocols. Furthermore, the multiphoton approach also provides resilience to high channel noises as shown in  \cite{jha2024, jha2024multi}. The initial idea of multiphoton protocols was to perform Quantum Secure Direct Communication, i.e., sharing messages directly without the need to establish secret keys   \cite{2005decoy, 2017modeling}.  The major example of multiphoton protocol is the \textit{three-stage} QKD proposed by Dr. Kak in 2006 \cite{Kak2006-3Stage}. There are, however, certain weaknesses associated with the multiphoton approach, such as the higher risk of a Man-in-the-middle attack with lesser detection chances than the single-photon approach. 
\end{enumerate}

We would now go over some of the different classes of QKD protocols. It is important to point out that all the protocol definitions are under the assumptions of an ideal environment, i.e., we have assumed no noise is present in the channel. The discussion of each of these protocols in a noisy environment is discussed at a later stage.

\begin{tcolorbox}[colback=gray!5,colframe=gray!40!black,title=Conceptual Overview: Uncertainty Based Protocols]
\begin{itemize}
  \item \textbf{Goal:} Force any eavesdropper into a guessing basis states for measurement using non-commuting bases.
  \item \textbf{Core Mechanism:} Randomly encode bits in one of two bases so wrong-basis measurements flip phase or bit values.
  \item \textbf{Loophole:} Practical photon sources sometimes emit multi-photon pulses, which makes them more susceptible to undetected PNS attacks. 
  \item \textbf{Key Fix:} Decoy-state methods interweave different intensities to detect tampering via statistical anomalies.
\end{itemize}
\end{tcolorbox}

\subsection{Bennett-Brassard 1984 (BB84) Protocol}

The BB84  \cite{bennett1984update} is the foundation for all Quantum Key Distribution protocols. This process is completed by establishing a secret key between Alice and Bob. The protocol starts with the preparation of qubits using two choices of bases. The preparation choices are listed in Table(\ref{tab1}).

\begin{table}[h!]
\centering
\begin{tabular}{|c|c|c|}
\hline
\textbf{Bit Value} & \textbf{Basis} & \textbf{Qubit State} \\ \hline
0 & Rectilinear ($+$) & $\vert 0 \rangle$ \\ \hline
1 & Rectilinear ($+$) & $\vert 1 \rangle$ \\ \hline
0 & Diagonal ($\times$) & $\vert + \rangle$\\ \hline
1 & Diagonal ($\times$) & $\vert - \rangle$ \\ \hline
\end{tabular}
\caption{Preparation choices in BB84 protocol}
\label{tab1}
\end{table}
The following are the entire steps of the BB84 protocol, 
\begin{enumerate}[label=(\roman*)]
    \item Alice prepares the qubits string for a random classical bit string using preparation choices as listed in Table(\ref{tab1}), and transmits these using the quantum channel.
    \item Bob randomly chooses a basis of measurement for each qubit. He keeps a record of both the qubit measured and the basis of choice.
    \item Bob declares his basis choice over the classical channel, and the qubits measured in different bases are discarded. 
    \item Alice and Bob then choose half of the remaining qubits to determine if an eavesdropper is present by using sifting. They discard these qubits used in sifting to get the final raw key.
    \item Once a raw key is established, Alice and Bob should end up with a secret key. They perform privacy amplification to reduce the amount of information leaked to eavesdroppers to a minimum. 
\end{enumerate}

\begin{tcolorbox}[colback=gray!5,colframe=gray!40!black,title=Conceptual Overview: Hybrid Architecture]
\begin{itemize}
  \item \textbf{Goal:} Transmit secret messages directly over quantum channels without pre-shared keys, ensuring eavesdropper detection at each stage.
  \item \textbf{Core Mechanism:} Based on the idea of double lock cryptography. 
  \item \textbf{Practical Loophole:} Trojan-horse and side-channel attacks can exploit the bidirectional nature of the protocol to inject or measure auxiliary photons.
  \item \textbf{Key Fix:} Introduce decoy-state pulses, optical isolators, and random timing to detect and thwart Trojan-horse attempts while monitoring channel losses.
\end{itemize}
\end{tcolorbox}
\subsection{Three-stage Protocol}

The three-stage protocol is based on a double-lock cryptography framework. This protocol was first proposed by Dr. Kak in 2006 and is multiphoton-resistant \cite{Kak2006-3Stage, chan2015multiphoton}. This protocol consists of three stages of encryption using unitary transformations on the qubit states. The following are the three stages for the execution of the protocol,
\begin{enumerate}[label=(\roman*)]
    \item Preparation: Alice prepares the quantum states in a pre-defined basis, i.e., the choice of basis is global knowledge. 
    \item Stage I: After preparing the qubits, Alice applies a unitary transformation, $U_A$, which is a rotation operator. The updated state can be written as, 
    \begin{equation}
        |\psi'\rangle = U_A|\psi\rangle.
        \label{S1}
    \end{equation}
    Alice transmits this updated state to Bob.
    \item Stage II: Bob receives the transmission states and applies his unitary transformation, $U_B$, without performing any measurements. The updated state is thus,
    \begin{equation}
        |\psi''\rangle = U_BU_A|\psi\rangle.
        \label{S2}
    \end{equation}
    Bob sends this updated state back to Alice.
    \item Stage III: Alice receives the updated qubits and removes her original unitary transformation, $U_A$, by applying an inverse of this transformation, $U_A^\dagger$. Given the nature of the unitary transformations, it is trivial to show that $[U_A, U_B]=0$. The updated qubit state is, 
    \begin{equation}
        |\psi'''\rangle = U_B|\psi\rangle.
        \label{S3}
    \end{equation}
    Alice transmits this state back to Bob, now, the final transmission stage.
    \item Measurement: Bob receives the final qubit states. He removes his unitary rotation operation and thus gets back the original qubit state. Bob now measures these qubits into the pre-defined basis and gets back classical bits. 
\end{enumerate}

It is essential to note that the three-stage protocol can be used as a quantum secure direct communication protocol, i.e., we can skip the key-distribution stage and directly transmit a message across a quantum channel. Therefore, it makes it important to study the variation of this protocol under several channel noises and several possible quantum attacks. We will look over these in later sections of this paper.

\begin{tcolorbox}[colback=gray!5,colframe=gray!40!black,title=Continuous Variable Protocols]
\begin{itemize}
  \item \textbf{Goal:} Leverage standard telecoms components to achieve high secret-key rates via Gaussian-modulated coherent states.  
  \item \textbf{Core Mechanism:} Alice encodes classical Gaussian variables onto the quadratures of coherent states; Bob performs homodyne or heterodyne detection to estimate those variables.  
  \item \textbf{Practical Loophole:} An adversary can manipulate or tamper with the local oscillator (LO) or inject excess noise to mask eavesdropping.  
  \item \textbf{Key Fix:} Real-time shot-noise monitoring, pilot-tone calibration, and receiver-side LO generation (or measurement-device-independent CV-QKD) detect or remove side-channel exploits.
\end{itemize}
\end{tcolorbox}

\subsection{Continuous Variable (CV) Protocols}
\label{Sec:CVQKD}
We would discuss the steps of CV QKD with Gaussian modulation. Similar to all QKD protocols, CV protocols also have three major steps: state preparation, transmission, and measurements. The following are the steps for CV QKD, 
\begin{enumerate}[label =(\roman*)]
    \item Preparation: Alice prepares states that are Gaussian modulation. Alice prepares a sequence of displaced coherent states, $|\alpha_1\rangle, |\alpha_2\rangle, \ldots |\alpha_N\rangle$ \cite{laudenbach2018}.
    \begin{equation}
        |\alpha_j\rangle= |q_j+ip_j\rangle,
        \label{cv-state}
    \end{equation}
    which obeys the following rules, 
    \begin{equation}
        \hat{a}|\alpha_j\rangle = \alpha_j|\alpha_j\rangle,
    \end{equation}
    where, 
    \begin{equation}
      \hat{a}=  \frac{1}{2}(\hat{q}+i\hat{p}).
    \end{equation}
    Each individual state has a mean photon number given as, 
    \begin{equation}
        \langle n_j\rangle = \langle\alpha_j|\hat{n}|\alpha_j\rangle = q_j^2+p_j^2.
    \end{equation}
    Once Alice has prepared the above-described sequence of coherent states, she transmits them to Bob. 

    \item Measurement: Based on the exact nature of the protocol, Bob can choose to measure either one quadrature component randomly at a time or can measure both quadrature components simultaneously (``no-switching protocol")  \cite{weedbrook2004}. Bob can use either of the measurement techniques (homodyne or heterodyne) to measure the eigenvalue of either or both quadrature operators. 

    \item Post-Measurement:
    \begin{itemize}
        \item \textbf{Sifting}: There are two possible preparations: choosing the basis for the construction of the protocol (similar to BB84). This requires sifting as it eliminates any uncorrelated qubits. In some forms of CV-QKD, both Alice and Bob use both bases simultaneously, which removes the need for sifting  \cite{laudenbach2018}.
        \item \textbf{Parameter Estimation}: After sifting, Alice and Bob share a subsection of the data that was sent to Bob and their corresponding measurements. This helps them estimate the amount of noise in the channel, which also helps to compute the mutual information, $I_{AB}$, and a bound on the eavesdropper's information, $\chi$   \cite{laudenbach2018}.
        \item \textbf{Information Reconciliation}: If the eavesdropper's information is found to be acceptable, i.e., $\chi < I_{AB} $, then Alice and Bob perform information reconciliation; otherwise, the protocol run is terminated. Alice and Bob now choose hash functions to further reduce the eavesdropper's information. Alice and Bob use these hash functions to find hash values. If hash values are the same, the keys are the same; otherwise keys must be discarded.
        \item \textbf{Privacy Amplification}: After getting the same keys, Alice and Bob would want to further reduce the eavesdropper's mutual information through this step. 
        
        \end{itemize}
\end{enumerate}

\begin{tcolorbox}[colback=gray!5,colframe=gray!40!black,title=Twin-Field (TF) Protocols]
\begin{itemize}
  \item \textbf{Goal:} Beat the repeater-less limit with key rate $\propto\sqrt{\eta}$ by interfering users’ faint, phase-encoded pulses at an untrusted relay.
  \item \textbf{Core Mechanism:} Alice/Bob phase-encode weak coherent states with decoy intensities and send to Charlie; a single click after a balanced beam splitter reveals the relative phase (Charlie only announces clicks/times).
  \item \textbf{Practical Loopholes:} Phase drift/locking demands, intensity mismatch, phase-reference leakage, and finite-key penalties at high loss.
  \item \textbf{Key Fix:} Pilot-tone or two-way phase lock with fiber-noise cancellation; discrete phase slicing and decoy estimation; composable finite-key analysis (MDI-style untrusted relay removes detector attacks).
\end{itemize}
\end{tcolorbox}

\subsection{Twin-field QKD Protocols}
\label{Sec:TFQKD}
{Twin-field QKD is one of the newer generations of QKD protocols. In this class of QKD protocols, Alice and Bob end faint, phase-related optical fields to an untrusted middle relay (often called Charlie). In this setup, only one photon needs to reach the center, the achievable secret-key rate scales with the square root of the channel transmittance, improving on direct transmission and surpassing the repeater-less limit at sufficient loss} \cite{lucamarini2018overcoming, curty2019simple}. {The detailed steps are as follows,}

\begin{enumerate}[label =(\roman*)]
    \item {Preparation: Alice and Bob randomly choose an intensity from a decoy set (e.g., signal plus two decoys) and a key bit, $b\in \{0,1\}$. }
    \begin{itemize}
        \item {Both Alice and Bob pick a global random phase (continuous) or a discrete phase-slice index (in phase-matching variants).}
        \item {Encode the bit by adding $0$ or $\pi$ phase to a weak coherent pulse, then transmit to Charlie over separate fibers.}
        \item {Phase randomization/slicing enables decoy-state parameter estimation.}
    \end{itemize}
    \item {Interference and measurement (by Charlie):}
    \begin{itemize}
        \item {Charlie interferes with the two inputs on a balanced beam splitter and monitors two single-photon detectors ($D_0$, $D_1$).}
        \item {Which detector clicks depends on the relative phase between Alice’s and Bob’s pulses (including their bit-dependent phase). Charlie publicly announces the detector clicks and their respective timing. }
    \end{itemize}
    \item {Post-measurement (public discussion and classical processing):}
    \begin{itemize}
        \item { Sifting and phase measurement: Alice and Bob reveal raw phase information (or phase-slice indices) and keep only matched events (the exact description depends on the particular variation of the TF QKD). Detector clicks are, thus, mapped to raw bits.}

        \item  {Decoy-state parameter estimation: Using the disclosed phase intensities, the single-photon yields are estimated, and the upper bound of information leakage is estimated. }

        \item {Information reconciliation and privacy amplification: mismatched bits are reconciled (for example, using the LDPC codes) and the keys are compressed to perform finite-key corrections to minimize Eve's mutual information. }
    \end{itemize}
    
\end{enumerate}

\subsection{Device Independent (DI) QKD}
\label{Sec:DIQKD}

{In this class of device-independent (DI) QKD, the devices used are treated as black-boxes and thus, certify secrecy solely from observed Bell-inequality violations, so security does not rely on modeling sources or detectors. The relay-free setup assumes only secure labs, free/random setting choices, and no unwanted leakage} \cite{le2025device}. {The steps are as follows,}
\begin{enumerate}[label =(\roman*)]
    \item {Preparation}:
    \begin{itemize}
        \item {Alice and Bob each have an uncharacterized device that receives classical inputs x and y (measurement settings) and returns outputs a and b (measurement results). }
        \item {Entanglement is distributed between the devices by any source (no restriction on the source: can be with either party, in the channel, or a completely untrusted source).}
        \item {Each round, every party chooses a measurement setting ($x,y$) from a predefined list and gives the provides the associated labels to the devices. For example, the pair ($x=0,y=0$) is reserved for key-generation, and other combinations are for testing and estimating any violations. }
    \end{itemize}
    \item {Measurement:}
    \begin{itemize}
        \item {For each round, the device outputs measurement results ($a$ and $b$) based on the label provided. The parties accumulate statistical results over multiple rounds. }
        \item {From these test rounds, the statistics are used to estimate a Bell parameter and quantum bit error rate (QBER). A loophole-free violation yields a device-independent bound on Eve's mutual information} \cite{hensen2015loophole, giustina2015significant}.
    \end{itemize}
    \item {Post Measurement:}
    \begin{itemize}
        \item {Sifting: keeps the rounds used for key-generation, discards the rest.}
        \item {Parameter Estimation: Test rounds are used to estimate the Bell-parameter $S$, and other parameters, such as error rates, are also calculated.}
        \item {Information reconciliation and privacy amplification: Correct mismatches in the raw keys, accounting for information leaked during error correction, and compressing the key to perform privacy amplification.}
    \end{itemize}
\end{enumerate}
\newcolumntype{Y}{>{\raggedright\arraybackslash}X}

\begin{table}[h!]
\small
\setlength\extrarowheight{2pt}
\centering
\caption{Comparative strengths and weaknesses of different QKD protocol classes.}
\begin{tabularx}{\linewidth}{|Y|Y|Y|}
\hline
\textbf{Protocol} & \textbf{Strengths} & \textbf{Weaknesses} \\ \hline

BB84\,/\,B92
  & \begin{itemize}[leftmargin=*,topsep=0pt,itemsep=0pt]
      \item Proven, information-theoretic security under ideal conditions.
      \item Conceptually simple and historically foundational.
    \end{itemize}
  & \begin{itemize}[leftmargin=*,topsep=0pt,itemsep=0pt]
      \item Requires true single-photon sources or decoy states to close the PNS loophole.
      \item Limited key rate over attenuated channels.
    \end{itemize}
  \\[4pt]\hline

Three-Stage QSDC
  & \begin{itemize}[leftmargin=*,topsep=0pt,itemsep=0pt]
      \item Direct secret message transmission without a prior key.
      \item Bidirectional eavesdropper checks.
    \end{itemize}
  & \begin{itemize}[leftmargin=*,topsep=0pt,itemsep=0pt]
      \item Vulnerable to Trojan-horse and side-channel injections.
      \item More complex hardware and timing requirements.
    \end{itemize}
  \\[4pt]\hline

Continuous-Variable QKD
  & \begin{itemize}[leftmargin=*,topsep=0pt,itemsep=0pt]
      \item High secret-key rates using standard telecom lasers and detectors.
      \item Naturally compatible with existing fiber infrastructure.
    \end{itemize}
  & \begin{itemize}[leftmargin=*,topsep=0pt,itemsep=0pt]
      \item LO side-channels (intensity/phase tampering).
      \item Generally limited to metropolitan distances without MDI extensions.
    \end{itemize}
  \\[4pt]\hline

{Twin-Field QKD}
  & \begin{itemize}[leftmargin=*,topsep=0pt,itemsep=0pt]
      \item {Secret-key rate scales $\sim\sqrt{\eta}$. It can surpass repeater-less bounds at sufficient loss.}
      \item {Un-trusted middle relay, and relay-detector side-channels largely neutralized.}
      \item {Works with weak coherent pulses and decoys. It is compatible with existing fiber as well.}
    \end{itemize}
  & \begin{itemize}[leftmargin=*,topsep=0pt,itemsep=0pt]
      \item {Requires precise phase stabilization/slicing and tight synchronization.}
      \item {Sensitive to reference-frame drift and intensity fluctuations.}
      \item {Finite-key and implementation overheads can reduce practical rates.}
    \end{itemize}
  \\[4pt]\hline

{Device-Independent QKD}
  & \begin{itemize}[leftmargin=*,topsep=0pt,itemsep=0pt]
      \item {Security certified from Bell-violation. No device modeling assumptions are used}.
      \item {Composable finite-key security via entropy accumulation.}
      \item {Tolerates internal side-channels if loopholes are closed.}
    \end{itemize}
  & \begin{itemize}[leftmargin=*,topsep=0pt,itemsep=0pt]
      \item {Requires high detection efficiency, low noise, and space-like separation. Short distances and modest rates today.}
      \item {Experimental complexity (loophole-free operation and strong RNG/isolation requirements)}
    \end{itemize}
  \\[4pt]\hline

\end{tabularx}
\label{Tab:CSQKD}
\end{table}

Having reviewed the principal classes of QKD protocols and distilled their core trade-offs (Table~\ref{Tab:CSQKD}), we now turn to the vulnerabilities those trade-offs invite. Each protocol’s unique challenges: multi-photon emissions in BB84, bidirectional exposure in three-stage QSDC, and local-oscillator side-channels in CV-QKD—create distinct entry points for Eve. In Sec.~\ref{Attacks}, we classify these attacks by source, channel, and detector, emphasizing which effects are hardest to distinguish from natural noise under finite data.

\section{Attacks on QKD}
\label{Attacks}
There are several possible attacks on the QKD, which became central to the discussions of practical QKD setups. The attacks on a quantum network can be broadly classified into three categories: Individual, Collective, and Coherent  \cite{wolf2021}. A lot of studies have been conducted on the most powerful attacks, i.e., Coherent attacks, which an eavesdropper can perform on a quantum channel, assuming no limitations on Eve's technological limits. We would look at some of the famous attacks that Eve can perform to compromise the integrity of quantum communication.

\begin{table}[!htpb]
\centering
\small
\caption{Broad Classification of QKD Attacks}
\renewcommand{\arraystretch}{1.2}
\begin{tabularx}{\textwidth}{|
  >{\centering\arraybackslash}p{0.18\textwidth}|
  >{\RaggedRight\arraybackslash}X|
  >{\centering\arraybackslash}p{0.18\textwidth}|
  >{\RaggedRight\arraybackslash}X|}
\hline
\textbf{Category} & \textbf{Examples} & \textbf{Primary Target} & \textbf{Typical Countermeasures} \\
\hline

Photon-Number Splitting Attacks
  & \begin{itemize}[leftmargin=*]
      \item Ideal PNS
      \item Non-ideal / performance-limited PNS
      \item Joint PNS (two-way)
      \item SPRINT-PNS
    \end{itemize}
  & Multi-photon source imperfections
  & Decoy-state protocols; detector monitoring; statistical tests \\[6pt] \hline

Trojan-Horse \& Side-Channels
  & \begin{itemize}[leftmargin=*]
      \item Back-scatter reflectometry (OTDR/OFDR)
      \item Delay-photon injection
      \item Gaussian / ML-crafted auxiliary pulses
    \end{itemize}
  & Transmitter \& receiver devices
  & Optical isolators; narrow-band filters; random timing; MDI/DI-QKD \\[6pt] \hline

Local-Oscillator Exploits (CV-QKD)
  & \begin{itemize}[leftmargin=*]
      \item LO intensity/phase manipulation
      \item Pilot-tone spoofing
    \end{itemize}
  & \begin{itemize}[leftmargin=*]
      \item Homodyne receivers
      \item Heterodyne receivers
    \end{itemize}
  & Local-LO generation; real-time shot-noise monitoring; pilot checks \\[6pt] \hline

Detector Attacks
  & \begin{itemize}[leftmargin=*]
      \item Detector blinding \& after-gating
      \item Timing attacks
      \item Wavelength-dependent attacks
    \end{itemize}
  & Single-photon detectors
  & MDI-QKD; watchdog detectors; active basis randomization \\[6pt] \hline

Jamming / DoS Attacks
  & \begin{itemize}[leftmargin=*]
      \item Polarization/Faraday jamming
      \item Arbitrarily-varying wiretap/jam channels (AVQC)
      \item Free-space reflective jamming
    \end{itemize}
  & Quantum channel / free-space links
  & Spatial filtering (SPAD arrays); frequency-hopping; error budgeting \\[6pt] \hline

Emerging ML-Guided Attacks
  & \begin{itemize}[leftmargin=*]
      \item Adaptive jamming via neural-network feedback
      \item Deep-learning side-channel inference
    \end{itemize}
  & All system components
  & ML-based anomaly detectors; provable ML-resilience frameworks \\ \hline
\end{tabularx}
\label{Tab:Taxonomy-Attacks}
\end{table}

\subsection{Photon Number Splitting (PNS) Attack}
\label{PNS}
\begin{figure*}[h!]
    \centering
    \includegraphics[width=\linewidth]{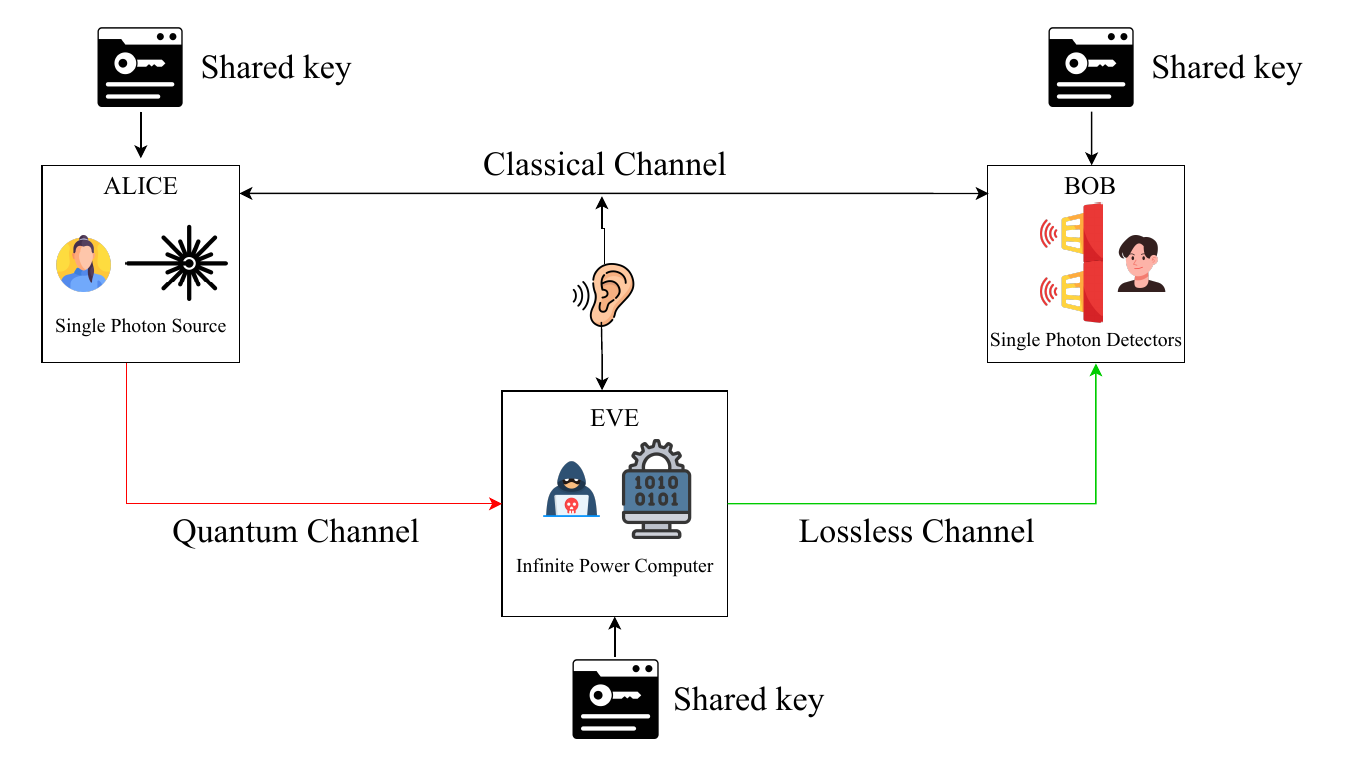}
    \caption{A schematic representation of a PNS attack conducted by an eavesdropper with infinite computational power. (Redrawn from  \cite{pnsieee}).}
    \label{fig:pns}
\end{figure*}

Photon Number Splitting (PNS) attack is theoretically one of the strongest attacks that can be performed by an eavesdropper on a quantum channel without introducing detectable errors \cite{brassard2000, lutkenhaus2000}. Fig. (\ref{fig:pns}) shows a schematic representation of the PNS attack conducted by an all-powerful Eve. The major steps in conducting a PNS attack are , 
\begin{enumerate}[label =(\roman*)]
    \item Eve starts by determining the number of photons, denoted by $n$, in each optical pulse that Alice generates. For any $n\geq 2$, Eve splits the multi-photon bursts and stores one photon internally. The number of photons in an optical pulse is measured using a specialized measurement in which a part of Alice's weak coherent pulse is projected and used without disturbing any of the encoded photons   \cite{lutkenhaus2000}. Else the pulse is blocked.
    \item For each pulse containing more than $n\geq 2$ photons, Eve sends the remaining $n-1$ photons to Bob via a lossless channel. 
    \item Eve keeps track of the classical communication between Alice and Bob to gain information about the preparation basis. Once basis measurements are known to Eve, she can measure any stored qubits by performing the attack. This allows Eve to obtain a copy of the secret key between the legitimate users. 
\end{enumerate}

\noindent Mathematically, a coherent state having a complex phase, $\alpha$, can be written as \cite{wolf2021}
\begin{equation}
    |\alpha\rangle = e^{\frac{|\alpha|^2}{2}}\sum_{n=0}^{\infty}\frac{\alpha^n}{\sqrt{n!}}|n\rangle,
    \label{coherent-phase}
\end{equation}
where $|n\rangle$ is the \textit{Fock} state representing a quantum state with a pulse containing $n$ photons. We can write the probability distribution of the number of photons by the Poisson distribution as, 
\begin{equation}
    P(n) = e^{-|\alpha|^2}\frac{|\alpha|^{2n}}{n!},
    \label{poisson}
\end{equation}
where $|\alpha|^2=\mu$ denotes the mean photon number of the signals. Thus, the state that Alice sends to Bob can be written as,
\begin{equation}
    \rho = e^{-\mu}\sum^{\infty}_{n=0}\frac{\mu^n}{n!}|n\rangle\langle n|,
    \label{state-pns}
\end{equation}
and the information is encoded in the polarization of these sent photons \cite{wolf2021}. We provided a brief overview of the ideal PNS attack. There are several variants of the PNS attack that have been theorized over the years.

\subsubsection{Related works on PNS-attack}
Below, we look at some of the work done concerning PNS attacks on different quantum key distribution protocols. 
\begin{enumerate}[label =(\roman*)]
    \item The work carried out by Mailloux et al \cite{pnsieee} simulates the PNS attack on the QKD system of the decoy state. The simulation study was broadly divided into two categories of PNS attacks, 
    \begin{itemize}
        \item Theoretically ideal PNS attack
        \item Non-ideal PNS attack, which accounts for performance limitations and several sources of real-life errors.
    \end{itemize}
    The authors present the results findings, which mainly focused on Eve's information gain and her detection probabilities for several PNS configurations. The authors show that in an ideal PNS attack scenario, Eve gains $100\%$ of the raw key bits \cite{pnsieee}. Moreover, they concluded that the decoy state protocol can detect PNS attack with high statistical confidence $P<0.001$ \cite{pnsieee}.

    \item Xiaoming et al. \cite{chen2022} propose a hypothesis‐testing approach to detect photon‐number splitting (PNS) attacks in decoy‐state MDI-QKD. They formulate the null hypothesis $H_0$,--\, ``no PNS attack”\,–\, and derive a test statistic $z$ assuming Gaussian fluctuations.  At the $5 \%$ significance level ($\lvert z\rvert>1.96$), they reject $H_0$ and discard affected pulses; otherwise, they accept $H_0$ and retain them.  Using realistic experimental parameters, they obtain $z=0.236$, yielding a Type I error probability below $5 \%$ and thus no attack detected under $H_0$.  However, this test is effective only against PNS attacks that significantly alter detection rates; subtler PNS strategies that leave overall statistics nearly unchanged may evade its detection.  

    \item Shang et al.\ \cite{mi2022joint} propose two joint‐PNS attacks on semi‐quantum QKD (SQKD), where a fully quantum Alice communicates with a classical Bob \cite{boyer2007quantum}. In both attacks, Eve exploits bidirectional channels to siphon off photons with minimal disturbance. In \textit{Attack 1}, she probabilistically blocks single‐photon pulses in both directions—achieving blocking rates of $88.7\%$ (forward) and $90.3\%$ (reverse) at a mean photon number $\mu = 0.20$. In \textit{Attack 2}, Eve first performs a nondestructive photon‐number measurement on the forward channel, then selectively blocks identifiable multiphoton pulses on the return, yielding parameters $\eta = 0.01$, $\mu = 0.02$, $p = 0.9905$, and $k = 1.036$. Both variants rely on high channel loss to remain covert; in low‐loss or error‐corrected SQKD implementations, their stealth and effectiveness are greatly reduced.

    \item The work presented by Wei-Tao Liu, et al., presents of principal experiment for a modified PNS attack on QKD network \cite{pnsaps}. This modified scheme is based on weak laser pulses. This modified approach does not require eavesdroppers to have the ability to isolate individual photons from multiphoton pulses. Instead, the eavesdropper uses a beamsplitter to split each pulse, retaining part of it and conducting a quantum non-demolition measurement to detect whether it contains photons. The information between Eve and Alice is calculated by using mutual information calculated as follows,  \cite{pnsaps}
    \begin{equation}
      \textstyle \small I^{\text{EA}} =  1+e^{\text{EA}}\log_2 e^{\text{EA}} + (1-e^{\text{EA}})\log_2(1-e^{\text{EA}}),
      \label{mutual-pns}
    \end{equation}
    where $e^{\text{EA}}$ is the error rate between Alice and Eve. The authors have shown that the mutual information between Eve and Alice can be very high (as high as $0.92$) while the QBER between Alice and Bob remains around the level indicating no attack ($6.5\%$) \cite{pnsaps}. Even though Alice or Bob does not discover the existence of Eve, Eve leaves a significant footprint in Bob's count. Under attack, the count rate is found to be quadratic; however, without attack, the loss of a real quantum channel should show a linear change \cite{pnsaps}. This can thus give the legitimate users an idea about possible eavesdropping attacks on the quantum channel! The key rate does show to be lower than expected. This is associated with several errors, especially the count rate due to the sensitivity of the key rate to the count rate \cite{pnsaps}.

    \item The work presented by Ashkenazy et al. proposes a \textit{SPRINT}-PNS attack \cite{PNS2024}. The authors present both theoretical and experimental results for the Sprint-PNS attack setup. The SPRINT mechanism occurs in three-level systems in the $\Lambda$ configuration, in which each transition is coupled to a separate photonic mode \cite{sprint}. In this description, the entire system is described in the form of a combined state, consisting of the early and late modes of transmitted ($T$) and reflected ($R$) spatial modes and the SPRINT state ($S$) \cite{PNS2024}. The modified QBER for the SPRINT-PNS attack is shown to have a linear approximation for large mean photon numbers, that is, $\mu$. For an eavesdropper without quantum memory, the authors showed that Eve can gain over $60\%$ of the information for a photon burst with $\mu \geq 1.5$, for different values of $\zeta$, i.e., the ratio of bits for which Eve couples the SPRINT \cite{PNS2024}. The results in this paper imply that an almost ideal PNS attack is possible with current technologies \cite{PNS2024}.

\end{enumerate} 
In ideal PNS attacks, Eve can recover the entire raw key without detection, and practical variants achieve over $90\%$ information gain at unchanged QBER. Decoy‐state and hypothesis‐testing defenses detect standard PNS with high confidence, but joint‐channel and memory‐assisted attacks evade detection under typical loss profiles. Experimental implementations using beam splitters and quantum non-demolition measurements confirm these vulnerabilities. Future work should aim to embed PNS detection into finite‐key analyses, design adaptive decoy sequences against advanced attacks, and integrate photon‐number‐resolving detectors with quantum error correction to limit Eve’s advantage in realistic channels.  

\subsection{Trojan Horse Attack}
\label{Sec:trojan}
\begin{figure*}[h!]
    \centering
    \includegraphics[width=0.75\linewidth]{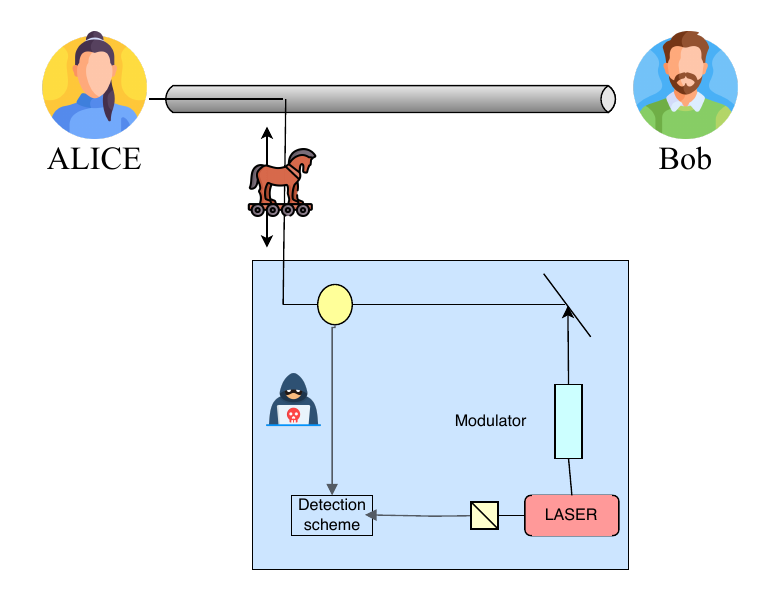}
    \caption{Schematic representation of the principle behind the Trojan horse attack. Eve occupies a section of the quantum channel (i.e., spatial, temporal, and frequency modes) to access Alice's apparatus. Eve uses an additional light source, modulates it, and then analyzes the backscattered signal with a detector. She can use features specific to her auxiliary source, such as its phase, in her detection configuration. Eve might occasionally have to intercept a portion of the valid signal to improve the quantum channel and make up for the loss. (Re-drawn from  \cite{gisin2006})}
    \label{fig:trojan}
\end{figure*}

Trojan horse attacks are one of the most powerful attacks on a QKD network. The term is taken from classical cryptography; however, in quantum Trojan-Horse attacks, Alice and Bob do not receive or accept any seemingly safe objects from the eavesdropper \cite{Gisin2002}. The main idea behind the Trojan horse attack is that the quantum channel can potentially allow the eavesdropper access to the legitimate user's apparatuses. Eve can exploit this by sending light pulses into Alice's or Bob's apparatus during the duration that the quantum channel is in use \cite{gisin2006}. Fig. (\ref{fig:trojan}) shows a schematic diagram representing the principle behind a general Trojan Horse attack on the quantum communication setup. 

To reduce the possibility of a Trojan horse attack, the system should be modified in the following way,  \cite{gisin2006}
\begin{itemize}
    \item The use of light filters such that only lights at appropriate wavelengths are allowed. 
    \item The encoding mechanism, i.e., the optical components, should only be active in short periods (the phase modulators are active only when qubits are being encoded).
    \item Bounding the level of the reflected light to keep an account of Eve exploiting it. 
\end{itemize}
In the fundamental work by Gisin et al., the authors presented the theoretical foundation for Eve's potential information gain while performing a Trojan Horse attack. To establish this, Eve has to start by distinguishing between two weak coherent states, $\ket{\alpha}\otimes \ket{0}$ and $\ket{0}\otimes \ket{\alpha}$. The measurement that would maximize the information gained by Eve can be utilized from  \cite{peres1997quantum}, and can thus be written as, 
\begin{equation}
    I^{\text{Trojan}}_{\text{Eve}}(|\alpha^2|) = 1-H(p),
    \label{info-trojan}
\end{equation}
where $H$ is the binary entropy and $p$ can be written as,
\begin{equation}
    p \approx \frac{1+\sqrt{2}|\alpha|}{2}.
    \label{eq:probab}
\end{equation}
Thus, eq(\ref{info-trojan}) can be approximated as, 
\begin{equation}
    I^{\text{Trojan}}_{\text{Eve}}(|\alpha^2|) = \frac{1}{\ln (2)}|\alpha|^2 + O(|\alpha|^4).
\end{equation}
This is noted to be larger than the probability of a weak-pulse being non-empty, 
\begin{equation}
    P(\text{non-empty}) = 1-\exp (-|\alpha|^2)\approx |\alpha|^2.
    \label{non-empty-weak}
\end{equation}
The author points out that this arises due to the assumption that Eve holds a coherent state, i.e., a phase reference relative to which $\alpha$ is defined. This is later used to reduce Eve's information gain to $|\alpha|^2$ \cite{gisin2006}.
\subsubsection{Related works to Trojan Horse attacks}
In this section, we review some of the works related to Trojan horse attacks on quantum communication. 
\begin{enumerate}[label =(\roman*)]
    \item Gisin et al.\ \cite{gisin2006} experimentally demonstrate Trojan‐horse attacks on QKD by injecting weak light pulses and analyzing the back‐scattered signal using both Optical Time‐Domain Reflectometry (OTDR) and Optical Frequency‐Domain Reflectometry (OFDR). They find that allowing just $0.1$ back‐scattered photons enables Eve to gain up to $ 0.135 $ bits per qubit, and that randomizing Alice’s phase settings reduces Eve’s information by a factor of approximately $1.44$—thereby lowering the required privacy amplification. Their analysis, however, assumes auxiliary detectors can instantly flag anomalous intensities and proposes ``optical fuses” that, while theoretically sound, are not yet practical with today’s technology.

    \item Yang et al.\ \cite{yang2015trojan} propose a Trojan‐horse attack on the BKM07 semi‐quantum key distribution protocol, targeting its classical Bob implementation. After Alice sends qubits prepared in the $X$ or $Z$ basis, Eve intercepts the pulses and injects a fake photon at the same wavelength but with a shorter delay than Bob’s detection window, allowing undetected extraction during the SIFT phase. To counter both delay‐photon injection and photon‐number‐splitting attacks, Bob replaces the ideal yet impractical photon‐number splitter with a wavelength filter and a $50/50$ photon beam splitter, blocking unauthorized pulses while preserving signal integrity. This modification secures BKM07 against the described Trojan‐horse variants, but extending these defenses to other SQKD schemes remains an important avenue for future work.  

    \item Suschchev et al.\ \cite{sushchev2024trojan} experimentally analyze Trojan‐horse attacks on BB84 QKD systems, deriving fidelity bounds for both pure and mixed back‐scattered states and assessing countermeasure performance via OTDR and spectral transmittance measurements.  The transmittance at wavelength $\lambda$ is given by

    \begin{equation}
    T[\mathrm{dB}]
    =10\log_{10}\frac{N - N_\text{dark}}{N_\text{ref} - N_\text{dark}}
    + A_\text{ref}[\mathrm{dB}] - A[\mathrm{dB}],
    \end{equation}
    
    where $N$, $N_\text{ref}$, and $N_\text{dark}$ denote the device, reference, and dark counts, and $A_\text{ref}$, $A$ the corresponding attenuation.  They report fidelities approaching $100\%$.  At $1550\,$nm, Eve’s mean photon number per pulse is $\mu_\text{Eve}=4\times10^{-16}$; at $1800\,$nm it rises to $9\times10^{-4}$, implying roughly $1\%$ potential key leakage.  This study is limited to the $1100\text{--}1800\,$nm range dictated by detector sensitivity; future work should investigate back‐flash radiation and radio‐frequency side channels from detectors and electronics.

    \item The work by Vinay et al. extends the analysis of Trojan-horse attacks by evaluating Gaussian attack states, and their impact on the secret key rate \cite{vinay2018extended}. Eve's Gaussian attack states exploit side channels in quantum key distribution (QKD). These states are constructed as follows:
    \begin{equation}
        \rho = D(\alpha) S_2(\xi_E) |0\rangle \langle 0| S_2^\dagger(\xi_E) D^\dagger(\alpha),
    \end{equation}
    where \(D(\alpha)\) is the displacement operator and \(S_2(\xi_E)\) is the two-mode squeeze operator. Attenuation and thermal noise are modeled with parameters \(\eta\) and \(\mu_T\), reducing Eve's information gain. Fidelity \(F(\rho_1, \rho_2)\) quantifies distinguishable, impacting the secret key rate:
    \begin{equation}
        K = R[1 - 2H_2(\epsilon) - \Delta].
    \end{equation}
    Coherent states are optimal among Gaussian attacks, even under noise, emphasizing the importance of attenuation in securing QKD systems. The main idea was to define a bound of the key rates under Trojan horse attacks.

    Eve's optimal strategy in Gaussian attacks on quantum key distribution (QKD) involves using coherent states, which are most effective among Gaussian states. The presence of thermal noise reduces Eve's ability to extract information, thereby enhancing the security of the system. The distinguishable ($(\Delta)$) of states plays a critical role; if it surpasses a certain threshold, the secret key rate becomes zero, leaving the QKD system vulnerable. Moreover, the distinguishable for separable states is higher than for coherent-state attacks under thermal noise, emphasizing the importance of practical defenses like attenuation and noise to secure QKD against these advanced side-channel attacks. The work addresses some important research questions about the Trojan-horse attack; however, the work can further be extended to non-Gaussian states and include some practical imperfections in QKD applications rather than idealized attenuation defenses.

    \item Jain et al.\ \cite{jain2014risk} analyze wavelength‐dependent vulnerabilities in QKD components—optical isolators, circulators, and single‐photon avalanche diodes (SPADs)—using spectral transmittance measurements.  They demonstrate that finite isolation permits Eve to exploit specific bands (e.g.\ $1300\text{--}1700\,$nm) where attenuation is inadequate and propose multi‐layer defenses combining isolators, monitoring detectors, and optimized fiber selection.  Their key findings are:  
    \begin{enumerate}[leftmargin=*]
      \item A $30\,$dB reduction in isolation at $1300\,$nm compared to $1500\,$nm.
      \item Bright pulses exceeding $4\times10^6$ photons bypass all protections.
      \item SPAD afterpulsing at $1700\text{--}1800\,$nm is $10^3$ times lower than at $1550\,$nm, increasing side‐channel risk.
    \end{enumerate}
    This study assumes idealized detector and filter performance; real‐world imperfections in these countermeasures merit further investigation.  

\end{enumerate}

To sum up, Trojan-horse attacks pose a pervasive threat to both discrete‐ and continuous‐variable QKD implementations—from back-scatter reflectometry in BB84 \cite{gisin2006} and semi-quantum variants \cite{yang2015trojan} to Gaussian-state side channels \cite{vinay2018extended} and wavelength-dependent leaks in isolators and SPADs \cite{jain2014risk,sushchev2024trojan}.  While techniques such as narrow-band filtering, random phase modulation, decoy-state defenses, and SPAD arrays can significantly limit Eve’s information gain or detect intrusion, open challenges remain. Future work should aim to develop unified security models that incorporate non-Gaussian probes and realistic device imperfections, extend countermeasures to emerging platforms (e.g.\ integrated photonics and free-space links), and explore active anomaly-detection schemes to detect and mitigate possible quantum-Trojan horse attacks.

\subsection{Jamming Attack}
\label{Sec:jamming}
Similar to classical communication scenarios eavesdropper can perform jamming attacks to quantum communication. A jamming attack is a denial-of-service (DoS) strategy. The main idea is for the eavesdropper to introduce enough noise in the system to disrupt the quantum key distribution protocols or to reduce their efficiency. 

Eve's jamming signal can be represented as an additional noise term while writing the channel's density matrix. The modified channel state under Eve's injection can be written as, 
\begin{equation}
    \rho_{\text{channel}} = (1-p)|\psi\rangle\langle\psi|+p\rho_E,
    \label{eq:denschannel}
\end{equation}
where $\rho_E$ is the quantum jamming signal. The channel noise can be represented as, 
\begin{equation}
    \mathcal{E}(\rho) = (1-p)\rho + p\rho_E,
\end{equation}
where:
\begin{itemize}
    \item $\rho$ is the density matrix of the transmitted state.
    \item p is the probability of noise injection by Eve.
    \item $\rho_E$ represents Eve's jamming signal.
\end{itemize}
Based on this noise injection, the fidelity of the states would be significantly reduced. We can write the fidelity between the original stat, $|\psi\rangle$, and received state $\rho_{\text{channel}}$ as, 
\begin{equation}
    F = \langle\psi|\rho_{\text{channel}}|\psi\rangle = 1-p+p\cdot F(\psi, \rho_E).
    \label{fid}
\end{equation}
if we consider $\rho_E$ to be depolarizing channel, then eq(\ref{fid}) can be reduced to, 
\begin{equation}
    F = 1-\frac{p}{2},
\end{equation}\
where $p$ is the probability of the application of the injection noise by the eavesdropper. This is the basic mathematical setup of jamming attack in QKD system.

\subsubsection{Related works on Jamming attacks}
We will go over some of the works related to jamming attacks in quantum communications scenarios. 
\begin{enumerate}[label =(\roman*)]

    \item Daschner et al.\ \cite{daschner2019exploiting} demonstrate a jamming attack on BB84 QKD leveraging the Faraday effect. An external magnetic field $B_0$ rotates photon polarization by  
    \begin{equation}
      \beta = V\,B_0\,L,
      \label{faraday}
    \end{equation}
        where, 
    \begin{itemize}
        \item $V$: Verdet constant (material and wavelength dependent),
        \item $B_0$: Magnetic field strength,
        \item $L$: Length of the medium through which light propagates.
    \end{itemize}
    In standard telecom fiber ($\lambda=1550$\,nm, $V\approx0.53\,\mathrm{rad\,T^{-1}m^{-1}}$), this rotation degrades the Bell‐inequality parameter to  
    \begin{equation}
      S' = S\,\bigl|\cos(2\beta)\bigr|
      \label{bell-vio}
    \end{equation}
    and induces a quantum bit‐error rate of  
    \begin{equation}
      \mathrm{QBER}_{\mathrm{ind}} = 1 - \cos^2(\beta).
      \label{qber}
    \end{equation}
    They show that $\beta=12.4^\circ$ lowers $S$ from 2.37 to 2.15 (confidence drops from $4\sigma$ to $1.7\sigma$), and $\beta=16^\circ$ increases QBER by 7.6\%, effectively denying service. The attack is less potent in free‐space links (lower atmospheric $V$) and its success depends critically on fiber proximity and lack of magnetic shielding.

    \item Boche et al. ($2017$)\ \cite{boche2017classical} model an arbitrarily varying classical–quantum wiretap channel under simultaneous passive eavesdropping and active jamming as
    \begin{equation}
    \mathcal{W} = \bigl(\mathcal{X},\mathcal{S},\mathcal{H}_B,\mathcal{H}_E,\{W_s^B,W_s^E\}_{s\in\mathcal{S}}\bigr),
    \label{channel}
    \end{equation}
    where $\mathcal{X}$ is the input alphabet, $\mathcal{S}$ the jammer’s state set, and $W_s^B$ ($W_s^E$) the channel to Bob (Eve) when in state $s$.  The deterministic secrecy capacity is defined by
    \begin{equation}
    C_s = \max_{p(x)}\inf_{s\in\mathcal{S}}\bigl[I(X;B\mid s)-I(X;E\mid s)\bigr],
    \label{secrecy}
    \end{equation}
    where $I(X;B\mid s)$ and $I(X;E\mid s)$ denote the mutual information between the input $X$ and Bob’s or Eve’s output under state $s$.  A channel is \emph{symmetrizable} if there exists $q(s\mid x)$ such that
    \begin{equation}
    \sum_{s\in\mathcal{S}}q(s\mid x)\,W_s^B \;=\;\sum_{s\in\mathcal{S}}q(s\mid x')\,W_s^B
    \quad\forall\,x,x'\in\mathcal{X},
    \end{equation}
    in which case $C_s=0$; otherwise $C_s>0$.  They further prove super-activation: two channels each with $C_s=0$ can together yield $C_s>0$.  Limitations include the assumption of full knowledge of adversarial capabilities and the absence of practical simulation or implementation guidance.

    \item Boche et al. ($2019$)\ \cite{boche2019simultaneous} study simultaneous classical and quantum communication over compound and arbitrarily varying quantum channels (AVQCs), deriving robust coding schemes and capacity regions under active jamming.  An AVQC is defined as  
    \begin{equation}
    \mathcal{N}_{\mathrm{AVQC}}=\{\mathcal{N}_s: s\in\mathcal{S}\},
    \label{avqc_channel}
    \end{equation}
    where $\mathcal{S}$ is the jammer’s state set and $\mathcal{N}_s$ the channel for state $s$.  The deterministic secrecy capacity for combined classical‐quantum transmission is  
    \begin{equation}
    C_{\mathrm{det}}
    =\max_{\rho}\inf_{s\in\mathcal{S}}\bigl[I(\rho;B\mid s)-I(\rho;E\mid s)\bigr],
    \label{deterministic_capacity}
    \end{equation}
    with $I(\rho;B\mid s)$ and $I(\rho;E\mid s)$ the quantum mutual informations to Bob and Eve, respectively.  They also prove super‐activation:  
    \begin{equation}
    C\bigl(\mathcal{N}_1\otimes\mathcal{N}_2\bigr)>0
    \quad\text{even if}\quad
    C(\mathcal{N}_1)=C(\mathcal{N}_2)=0,
    \label{super_activation}
    \end{equation}
    showing that two zero‐capacity channels can jointly yield positive capacity.

    \item Simmons et al.\ \cite{simmons2023investigation} examine jamming threats to free‐space QKD, with a focus on satellite links. They consider two attack modes,

    \begin{itemize}[leftmargin=*]
      \item \textbf{Line‐of‐Sight (LoS)}: A ground‐based laser floods the receiver’s SPAD, overwhelming detection.  Experimentally, replacing a single‐pixel detector with a SPAD array increases the field of view and yields an improvement of approximately $20\,$dB in the signal‐to‐interference ratio (SIR).
      
      \item \textbf{Non‐Line‐of‐Sight (NLoS)}: Reflective jamming uses ground lasers to illuminate the satellite, which then scatters light back to the ground station.  Simulations show this saturates the LoS link budget, making QKD infeasible during satellite passes.
    \end{itemize}
    
    While SPAD arrays effectively mitigate direct blinding, countering reflective jamming remains an open challenge, as distinguishing legitimate quantum signals from reflected interference is difficult.

    \end{enumerate}
To review, the body of work on jamming attacks in QKD—from Faraday‐effect based polarization rotation \cite{daschner2019exploiting} to arbitrarily varying classical–quantum wiretap models \cite{boche2017classical,boche2019simultaneous} and free‐space satellite jamming \cite{simmons2023investigation}—demonstrates both the feasibility of denial‐of‐service strategies and the promise of countermeasures such as SPAD arrays, frequency‐domain defenses, and robust coding schemes. Yet several challenges remain.  Future research can aim to develop unified models that combine passive and active jamming under realistic hardware constraints, explore dynamic countermeasures like frequency hopping or adaptive nulling, and integrate quantum error correction codes to correct jamming‐induced logical errors.  In addition, hardware‐in‐the‐loop experiments and high‐dimensional encoding schemes could provide further resilience, while machine learning–based anomaly detection can also be explored as event detection for more sophisticated jamming attacks. In tab(\ref{tab:security}), we summarize all the attacks we reviewed in this Sec \ref{Attacks}.

\begin{table}[!htpb]
\renewcommand{\arraystretch}{1.5} 
\centering
\caption{Summary of Related works on several Attacks on QKD networks.}
\label{tab:attack_summary}
\begin{tabular}{|l|l|l|p{5cm}|}
\hline
\textbf{Type of Attack}       & \textbf{Year} & \textbf{Reference}       & \textbf{Summary} \\ \hline

\multirow{4}{*}{PNS Attack} 
    & 2000 & Mailloux et al \cite{pnsieee}.         & Simulates PNS attacks on decoy-state QKD, analyzing Eve's information gain and detection probabilities. \\ \cline{2-4}
    & 2022 & Xiaoming et al \cite{chen2022}.        & Uses hypothesis testing to detect weaker forms of PNS attacks on decoy-state QKD. \\ \cline{2-4}
    & 2024 & Liu et al \cite{pnsaps}.               & Demonstrates a beamsplitter-based PNS attack without isolating individual photons, showing high mutual information. \\ \cline{2-4}
    & 2024 & Ashkenazy et al \cite{PNS2024}.        & Proposes SPRINT-PNS attack with experimental results showing high Eve information gain for large $\mu$. \\ \hline

\multirow{3}{*}{Trojan Horse Attack} 
    & 2006 & Gisin et al \cite{gisin2006}.          & Analyzes back-scattered signals to gain information and proposes practical countermeasures like optical fuses. \\ \cline{2-4}
    & 2015 & Yang et al \cite{yang2015trojan}.      & Proposes a Trojan horse attack and mitigates it using photon beam splitters and wavelength filters. \\ \cline{2-4}
    & 2024 & Sushchev et al \cite{sushchev2024trojan}. & Defines bounds on fidelity using OTDR and spectral transmittance for Trojan Horse attacks. \\ \hline

\multirow{4}{*}{Jamming Attack} 
    & 2017 & Boche et al \cite{boche2017classical}. & Analyzes secrecy capacity for channels under jamming and eavesdropping with super-activation effects. \\ \cline{2-4}
    & 2019 & Daschner et al \cite{daschner2019exploiting}. & Exploits the Faraday effect to disrupt QKD, increasing QBER and halting key exchange. \\ \cline{2-4}
    & 2019 & Boche et al \cite{boche2019simultaneous}. & Derives capacity regions for compound and AVQCs, demonstrating simultaneous classical-quantum transmission. \\ \cline{2-4}
    & 2023 & Simmons et al \cite{simmons2023investigation}. & Explores line-of-sight and reflective jamming on satellite QKD, improving SIR with SPAD arrays. \\ \hline

\end{tabular}
\label{tab:security}
\end{table}

{Because many of these attacks are statistically indistinguishable from background noise at realistic block sizes, asymptotic analyses alone are insufficient.} In Sec.~\ref{Security-analysis} {we map these threats to unconditional-, finite-key-, and composable-security frameworks. We also highlight the role of QECC in keeping error budgets within those bounds.
}

\section{Security Analysis of QKD}
\label{Security-analysis}
We now review the security of Quantum Key Distribution systems by looking over different classes of security proofs. 

\subsection{Security Paradigms in QKD}
As mentioned earlier, QKD provides security by leveraging the principles of quantum mechanics. The security paradigms in QKD are broadly classified as unconditional security, finite-key security, and composable security.

QKD exploits inherently quantum phenomena—such as non-orthogonal state encoding and the no-cloning theorem—to generate shared keys between Alice and Bob. This provides information-theoretic security rather than security relying on hard mathematical problems \cite{Gisin2002,Scarani2009}. Since BB84’s formulation, several rigorous proofs have established its unconditional security under ideal assumptions, in stark contrast to classical schemes (e.g.\ RSA) whose safety depends on computational complexity \cite{Scarani2009,yuen2012unconditional}. Security analyses further distinguish raw security—Eve’s maximum success probability during key generation—from Known-Plaintext Attack (KPA) security, which bounds her advantage once part of the key is revealed during message exchange \cite{yuen2010fundamental,yuen2012unconditional}.

\begin{enumerate}
 \item As described in  \cite{yuen2012unconditional}, UCS implies that the adversary's probability of successfully estimating a subset \( K^* \) of the key (or the entire key \( K \)) is negligible. For raw security, this probability satisfies:
\begin{equation}
    p_1(K^*) \leq 2^{-|K^*|} + \epsilon',
\end{equation}
where \( \epsilon' \) is a small bound. Under a Known-Plaintext Attack (KPA), the probability is bounded as:
\begin{equation}
    p_1(K_2^* | K_1 = k_1) \leq 2^{-|K_2^*|} + \epsilon'',
\end{equation}
where \( K_1 \) is the known segment, \( K_2 \) is the remaining subset, and \( \epsilon'' \) is another small bound. Trace distance quantifies how close the actual key distribution \( \rho_{KE} \) is to the ideal uniform key distribution \( \rho_U \), and is defined as,
\begin{equation}
    d = \frac{1}{2} \| \rho_{KE} - \rho_U \otimes \rho_E \|_1,
\end{equation}
where \( \| \cdot \|_1 \) is the trace norm. For \( d \leq \epsilon \), the adversary's success probability is bounded by:
\begin{equation}
    p_1(K) \leq \frac{1}{2^n} + \epsilon,
\end{equation}
and for KPA, the averaged success probability satisfies:
\begin{equation}
    p_1(K_2^* | K_1) \leq 2^{-|K_2^*|} + \epsilon.
\end{equation}

Bit Error Rate (BER) measures the probability of the adversary correctly guessing individual bits of \( K^* \), even when the sequence \( K^* \) is incorrect. It is bounded by Fano's inequality as:
\begin{equation}
    n H(p_b) \geq H(K) - I_{\text{ac}},
\end{equation}
where \( p_b \) is the bit error probability, \( H(\cdot) \) is entropy, and \( I_{\text{ac}} \) is accessible information.

The trace distance criterion \( d \leq \epsilon \) provides useful bounds for adversary success probabilities but does not guarantee perfect security for partial keys under KPA. Practical implementations require significantly smaller \( d \) values to ensure security and avoid breaches. Both sequence error probabilities and BER must be rigorously bounded to achieve UCS. The analysis highlights that widely accepted trace distance criteria may overestimate the security level in QKD, especially in finite-key and KPA scenarios. Achieving UCS requires careful mathematical bounds and significantly smaller trace distance values.

\item The work by Boileau et al. presents the principles behind the proof of unconditional security of the three-state quantum key distribution (QKD) protocol. The protocol, also called the \textit{trine spherical code protocol} (PBC00), demonstrates security up to a bit error rate of \( 9.81\% \), making it comparable to the BB84 protocol but with a reduced number of states \cite{boileau2005unconditional}. The PBC00 protocol uses three states that form an equilateral triangle on the X-Z plane of the Bloch sphere. These states are defined as:
\begin{align}
    |1\rangle &= \frac{1}{2} |0_x\rangle + \frac{\sqrt{3}}{2} |1_x\rangle, \nonumber\\
    |2\rangle &= \frac{1}{2} |0_x\rangle - \frac{\sqrt{3}}{2} |1_x\rangle, \nonumber\\
    |3\rangle &= |1_x\rangle.
\end{align}

where \( |0_x\rangle \) and \( |1_x\rangle \) form the X-basis. Bob performs a measurement using a Positive Operator-Valued Measure (POVM) with elements:
\begin{align}
    E_1 &= \frac{2}{3} |\tilde{1}\rangle \langle \tilde{1}|, \nonumber\\
    E_2 &= \frac{2}{3} |\tilde{2}\rangle \langle \tilde{2}|, \nonumber\\
    E_3 &= \frac{2}{3} |\tilde{3}\rangle \langle \tilde{3}|,
\end{align}
where \( |\tilde{i}\rangle \) are the states orthogonal to Alice's original states.

The proof of security is established by relating the PBC00 protocol to a QKD protocol based on entanglement distillation protocols (EDP) using Calderbank-Shor-Steane (CSS) codes. Alice prepares entangled pairs in the state,
\begin{equation}
    |\Phi\rangle = \frac{1}{\sqrt{2}} \left( |0_z\rangle_A |1\rangle_B + |1_z\rangle_A |2\rangle_B \right),
\end{equation}
where Bob applies a rotation \( R_y(2b\pi/3) \) on his qubit depending on Alice's trit string \( r \). A local filtering operation, represented by the Kraus operator:
\begin{equation}
    F = |0_x\rangle \langle 0_x| + \frac{1}{\sqrt{3}} |1_x\rangle \langle 1_x|,
\end{equation}
distills maximally entangled pairs \( |\Phi^+\rangle \) when successful.

The phase error rate \( e_{\text{phase}} \) is deduced from the bit error rate \( e_{\text{bit}} \) using Azuma's inequality, which bounds the probability of large deviations in correlated random variables. The phase error rate is shown to be:
\begin{equation}
    e_{\text{phase}} = \frac{5}{4} e_{\text{bit}}.
\end{equation}
By measuring test bits, Alice and Bob estimate the bit error rate \( e_{\text{bit}} \), which determines the phase error rate. Using privacy amplification and error correction, the achievable key rate is:
\begin{equation}
    R = p_{\text{conc}} \left[ 1 - h(e_{\text{bit}}) - h\left(\frac{5}{4} e_{\text{bit}} \right) \right],
\end{equation}
where $p_{\text{conc}}$ is the probability of conclusive events, and \( h(x) \) is the binary entropy function. The protocol remains secure up to a bit error rate of $e_{\text{bit}} = 9.81\%$, where the key rate reaches zero. The proof applies equally to the modified R04 protocol, where the error rate is estimated from the number of inconclusive results instead of using test bits. By modifying the protocol so that Alice chooses a random basis only after Bob confirms the receipt of a photon, the inconclusive events provide an equivalent estimate of the error rate.

\item Finite key analysis is another important aspect of the security analysis of QKD protocols. While the traditional security proofs of QKD assume infinite key length (i.e., an asymptotic scenario), for practical purposes, only a finite length key can be established. Thus, we need to do finite key analysis to address the statistical fluctuations and uncertainties that arise when working with limited resources. The work presented by Bunandar et al. presents a numerical approach for calculating finite-key secret key rates in Quantum Key Distribution (QKD), addressing practical constraints when only a limited number of quantum signals are exchanged  \cite{bunandar2020numerical}. The authors develop two semi-definite programs (SDPs) to compute the finite-key rates using smooth min-entropy and relative entropy, ensuring composable security. The secure key length is quantified using the smooth min-entropy \( H_{\text{min}}^{\epsilon} \), approximated by the quantum relative entropy \( H_{\text{rel}} \) between the state \( \rho \) (signal state) and a reference state \( \sigma \):
\begin{equation}
    H_{\text{min}}^{\epsilon}(\rho) \approx H_{\text{rel}}(\rho \| \sigma).
\end{equation}

The key rate \( R \) in the finite regime is bounded as:
\begin{equation}
    R = p_{\text{succ}} \left[ 1 - h(e_{\text{bit}}) - h(e_{\text{phase}}) \right],
\end{equation}
where \( p_{\text{succ}} \) is the probability of conclusive events, \( e_{\text{bit}} \) is the bit error rate, and \( e_{\text{phase}} \) is the phase error rate. The framework is applied to:
\begin{itemize}
    \item \textbf{BB84 with Detector Asymmetries}: Incorporates unequal detector efficiencies.
    \item \textbf{B92 Protocol}: Handles asymmetric non-orthogonal states.
    \item \textbf{Twin-Field QKD}: Achieves key rates that surpass the PLOB bound, a fundamental limit for repeaterless communication.
\end{itemize}

By solving the SDPs using convex optimization, the approach provides reliable, numerically optimized key rates under finite resources. This method improves upon analytical bounds and is extensible to scenarios with decoy states and weak coherent pulses, offering a flexible and practical tool for real-world QKD implementations. This work does not focus on coherent attacks (more general strategies) are not fully addressed in the current framework. 

\item {Universal composable security} ensures that QKD remains secure even when the generated keys are reused or combined with other cryptographic protocols. The work by Ben-Or et al. introduces a composable security framework for QKD, proving robustness against general quantum attacks, including delayed measurements and joint operations  \cite{ben2005universal}.  

The security criterion is defined as the indistinguishability between the actual QKD protocol \( \kappa \) and an ideal key generation protocol \( \kappa_I \):
\begin{equation}
    \left\| \rho_{\text{qkd}} - \rho_{\text{ideal}} \right\|_1 \leq \epsilon,
\end{equation}
where \( \rho_{\text{qkd}} \) is the real key state, \( \rho_{\text{ideal}} \) is the perfect key state, and \( \epsilon \) quantifies Eve's advantage.

The framework relies on two main conditions:
\begin{itemize}
    \item \textbf{Equality-and-Uniformity}: The shared key is both identical and uniformly random:
    \begin{equation}
        \small
        \sum_m \Pr(M = m) \| p^{(m)}_{\text{ideal}} - p^{(m)}_{\text{qkd}} \|_1 \leq \mu_1.
    \end{equation}
    \item \textbf{Privacy Condition}: Limits Eve’s knowledge about the key using the trace distance bound.
\end{itemize}

The authors further use the \textit{singlet fidelity} \( F \) to bound the difference between the real and ideal key states,
\begin{equation}
    \small
    \frac{1}{2} \left\| \rho_{\text{qkd}} - \rho_{\text{ideal}} \right\|_1 \leq \sqrt{1 - F}.
\end{equation}

The framework establishes that protocols like {BB84} satisfy composable security, ensuring that the generated keys remain secure and usable, even in adversarial scenarios or when combined with other cryptographic systems. This provides a rigorous and general mathematical foundation for QKD security. {Finally, to address implementation side-channels (e.g., detector control, LO tampering) that ideal proofs do not model, measurement-device-independent and device-independent variants certify secrecy from either an untrusted relay or a Bell violation, thus relocating trust away from the very components exploited in} Sec.~\ref{Attacks}.

\end{enumerate}

\subsection{Security Proofs}
\label{Sec:Security-proof}

\begin{enumerate}
    \item \textbf{Lo-Chau Security Proof}:  In this framework, Alice and Bob aim to distill near-perfect \textit{Einstein-Podolsky-Rosen (EPR)} pairs from noisy entangled states shared over an imperfect quantum channel \cite{lo1999unconditional}. These imperfections arise from channel noise or eavesdropping attempts by Eve. Alice and Bob start by sharing $m$ Bell states, ideally described as,
    \begin{equation}
        |\Phi^+\rangle = \frac{1}{\sqrt{2}} (|00\rangle + |11\rangle).
    \end{equation}
    However, due to imperfections, the actual shared state deviates from this form and may be entangled with Eve's system. To recover near-perfect entanglement, Alice and Bob apply quantum error correction as part of the entanglement distillation process. The fidelity \( F \) of the shared state \( \rho_{AB} \) to the ideal EPR pair \( |\Phi^+\rangle^{\otimes m} \) is then bounded as,  \cite{fung2010practical}
    \begin{equation}
        F(\rho_{ABE}, |\Phi^+\rangle^{\otimes m} \otimes \rho_E) \geq 1 - \epsilon_f,
    \end{equation}
    where \( \epsilon_f \) is the failure probability of the distillation process.
    
    After distillation, Alice and Bob measure the purified entangled states in the \( Z \)-basis to obtain a shared classical key. The trace distance between the actual key state \( \rho_{ABE} \) and the ideal key state \( \rho_{\text{ideal}} \) is bounded as,  \cite{xu2020secure}
    \begin{equation}
        \frac{1}{2} \|\rho_{ABE} - \rho_{\text{ideal}} \|_1 \leq \sqrt{\epsilon_f (2 - \epsilon_f)}.
    \end{equation}
    This trace distance ensures composable security, meaning that the key remains secure even if it is reused in other cryptographic protocols.
    
    The security analysis also involves estimating the bit error rate, \( e_b \) (errors in the \( Z \)-basis) and the phase error rate, \( e_p \) (errors in the \( X \)-basis). These error rates are defined as,

    \begin{align}
        e_b &= \frac{\text{No. of bit errors}}{N}, \nonumber\\
        e_p &= \frac{\text{No. of phase errors}}{N}.
    \end{align}
    By reducing these errors through quantum error correction, Alice and Bob ensure that the shared key is both accurate and secure. The Lo-Chau security proof is significant because it explicitly connects QKD security to entanglement distillation. This approach guarantees security against the most general form of eavesdropping, known as coherent attacks, where Eve can interact with all quantum signals and delay her measurements. By purifying entangled states, Alice and Bob ensure that Eve’s information about the final key is negligible, thereby establishing the unconditional security of QKD. However, as pointed out in  \cite{xu2020secure}, the entanglement distillation may fail with a low probability in real-life scenarios where the data size is finite.

    \item \textbf{Preskill-Shor Security Proof (BB84)}: This proof merges the entanglement distillation and error correction mechanisms to establish the security of the QKD \cite{xu2020secure, shor2000simple}. The security proof begins by interpreting BB84 as an entanglement-based protocol. Alice prepares entangled pairs of qubits in the state, 
    \begin{equation}
        |\Phi^+\rangle = \frac{1}{\sqrt{2}} \left( |00\rangle + |11\rangle \right),
    \end{equation}
    and sends one qubit to Bob through a noisy quantum channel. Due to errors caused by noise or Eve's interference, the shared state deviates from \( |\Phi^+\rangle \), and Alice and Bob aim to recover the entanglement using quantum error correction.
    
    To ensure security, Alice and Bob perform error correction and privacy amplification. Error correction reduces the bit error rate \( e_b \), which measures the probability of errors in the \( Z \)-basis:
    \begin{align}
        e_b &= \frac{\text{Number of bit errors}}{N}.
    \end{align}
    They also estimate the phase error rate \( e_p \), corresponding to errors in the \( X \)-basis:
    \begin{align}
        e_p &= \frac{\text{Number of phase errors}}{N}.
    \end{align}
    
    The key insight of the Shor-Preskill proof is that quantum error correction codes, specifically \textit{Calderbank-Shor-Steane (CSS) codes}, can simultaneously correct bit errors and phase errors. Alice and Bob effectively distill near-perfect entanglement from the noisy shared state by performing error correction. The fidelity of the corrected state to the ideal \( |\Phi^+\rangle \) is bounded as,
    \begin{equation}
        F \geq 1 - \epsilon,
    \end{equation}
    where \( \epsilon \) is the failure probability of error correction. After error correction, Alice and Bob apply privacy amplification to remove any information Eve may have gained during her interaction with the quantum channel. The final key rate \( R \) is given by,
    \begin{equation}
        R = 1 - h(e_b) - h(e_p),
    \end{equation}
    where \( h(x) \) is the binary entropy function, and \( e_b \) and \( e_p \) are the bit and phase error rates, respectively. The Shor-Preskill proof concludes that BB84 is secure if the bit error rate and phase error rate satisfy the following relation,
    \begin{equation}
        h(e_b) + h(e_p) < 1.
    \end{equation}
    This result ensures that Alice and Bob can generate a shared key that is both correct and private, even under the most general attacks by an eavesdropper. The Shor-Preskill proof is significant because it simplifies earlier, more complex security proofs. {Furthermore, the proof’s use of CSS codes makes explicit the mapping between observed bit or phase errors and correctable syndromes, thus, precisely the interface to the quantum error-correction machinery discussed afterwards.}

    \item \textbf{Composable Security Proof for CV Protocol}: The security proof begins by considering a reverse-reconciliation entanglement-based (EB) protocol \cite{leverrier2015composable}. Alice prepares \( 2n \) two-mode squeezed vacuum states and sends half of each state to Bob through a quantum channel \cite{xu2020secure}. Bob measures his received modes using heterodyne detection, and the shared quantum state is analyzed for parameter estimation. 

    The key length \( l \) in finite-size regimes is derived using the following bound,  \cite{leverrier2015composable}
    \begin{align}
        l \leq 2n \Big[ 
            & 2 H_{\text{MLE}}(U) 
            - f\left( \Sigma_a^{\text{max}}, \Sigma_b^{\text{max}}, \Sigma_c^{\text{min}} \right) 
        \Big] \notag \\
        & - \text{leak}_{\text{EC}} 
        - \Delta_{\text{AEP}} 
        - \Delta_{\text{ent}} \notag \\
        & - 2 \log \left( \frac{1}{2\bar{\epsilon}} \right).
    \end{align}

    where \( H_{\text{MLE}}(U) \) is the maximum-likelihood entropy of the discretized measurement outcomes \( U \), and \( f(\Sigma_a^{\text{max}}, \Sigma_b^{\text{max}}, \Sigma_c^{\text{min}}) \) is the Holevo bound quantifying Eve's accessible information. The terms \( \Delta_{\text{AEP}} \) and \( \Delta_{\text{ent}} \) account for finite-size deviations and estimation errors.
    
    To evaluate the covariance matrix of the state shared between Alice and Bob, the protocol relies on parameter estimation (PE) after error correction. The estimated covariance matrix parameters \( \gamma_a \), \( \gamma_b \), and \( \gamma_c \) are computed as:
    \begin{align}
        \gamma_a &\approx \frac{1}{2n} \left( \|X\|^2 - 1 \right), \nonumber\\
        \gamma_b &\approx \frac{1}{2n} \left( \|Y\|^2 - 1 \right), \nonumber\\
        \gamma_c &\approx \frac{1}{2n} \langle X, Y \rangle.
    \end{align}
    
    The PE step ensures that the covariance matrix parameters lie within confidence bounds \( \Sigma_a^{\text{max}}, \Sigma_b^{\text{max}}, \Sigma_c^{\text{min}} \) with high probability, allowing the secure key rate to be derived.
    
    The protocol achieves security against general attacks using the postselection technique or de Finetti's theorem, which reduces general attacks to collective attacks. The resulting secret key rate for \( n \) exchanged signals is:
    \begin{equation}
        R = \frac{l}{2n} \approx 1 - h(e_{\text{bit}}) - h(e_{\text{phase}}),
    \end{equation}
    where \( e_{\text{bit}} \) and \( e_{\text{phase}} \) are the bit and phase error rates, respectively.
    
    The analysis confirms that Gaussian attacks are asymptotically optimal under the composable security framework, and the finite-size key rates converge to their asymptotic values for sufficiently large \( n \). This work highlights a rigorous finite-key analysis for practical CV QKD, ensuring composable security under realistic conditions.

    \item \textbf{Security Proof for Three-stage Protocol}: The work by Chan et al. presents a security analysis of the three-stage (as proposed in  \cite{Kak2006-3Stage}), multi-photon quantum key distribution (QKD) protocol \cite{chan2015multiphoton}. This protocol leverages multi-photon coherent states and double-lock cryptography, providing robustness against three types of attacks: intercept-resend (IR), photon number splitting (PNS), and man-in-the-middle (MIM) attacks \cite{bagan2005comprehensive, zhao2007experimental}.

    The protocol's operation involves Alice and Bob applying commutative unitary transformations \( U_A \) and \( U_B \) to a quantum state \( X \) during transmission. Eve’s ability to eavesdrop is analyzed through her error probability in determining polarization states. 

    \begin{enumerate}

        \item Error Probability \( P_e \) under Intercept-Resend (IR) and Photon Number Splitting (PNS) Attacks is given by integrating over the polarization states:
   \begin{align}
       P_e = \int_S P(N_1, \phi_1) P(N_2, \phi_2) \, d\phi_1 d\phi_2,
   \end{align}
   where,
   \begin{itemize}
       \item \( P(N_i, \phi_i) \) is the probability distribution for photons in polarization states \( \phi_1 \) and \( \phi_2 \),
       \item \( S \) is the integration domain corresponding to Eve's estimation error for the polarization angles.
   \end{itemize}
    
    \item For MIM attacks, Eve impersonates Alice and Bob simultaneously, extracting polarization angles with photon loss constraints. The error probability under MIM attacks is computed as,
   \begin{align}
       P_{\text{Auth, MIM}} &= \int_{|\phi_1 - \phi_2| > \pi/4} P(N_t, \phi) \, d\phi,
   \end{align}
   where \( |\phi_1 - \phi_2| > \pi/4 \) represents the condition under which Eve's attack causes an authentication failure. The normal operation probability without attacks is given by:
   \begin{align}
       P_{\text{Auth, normal}} &= \int_{|\phi| > \pi/4} P(N, \phi) \, d\phi.
   \end{align}

   \item The secure key rate \( K \) is expressed as:
   \begin{equation}
       K = R \left[ 1 - h(Q) \right] - f \cdot \Delta P_e,
   \end{equation}
   where:
   \begin{itemize}
       \item \( R \) is the raw key generation rate,
       \item \( Q \) is the quantum bit error rate (QBER),
       \item \( h(Q) \) is the binary entropy function: 
             \[
                 h(Q) = -Q \log_2 Q - (1-Q) \log_2 (1-Q),
             \]
       \item \( f \) is a scaling factor for the fraction of man-in-the-middle attacks,
       \item \( \Delta P_e \) represents Eve's induced error probability.
   \end{itemize}

    \end{enumerate}

    The authors conclude that the three-stage protocol remains resilient to IR, PNS, and MIM attacks, especially for coherent states with a mean photon number \( N > 1 \). However, additional photon loss due to multiple transmissions poses challenges for long-distance communication. The work highlights the protocol's robustness but acknowledges limitations, such as sensitivity to amplification attacks and practical implementation constraints. Future developments include addressing the effects of noiseless amplification and state discrimination attacks.
\end{enumerate}

\section{Quantum Error Correction Codes}
\label{QECC}
Quantum error correction (QEC) is mainly based on classical coding theory; however, several important things must be considered to make it quantum analogous. The discussion of QEC requires a brief outline of the sources of errors in quantum information processing, and thus, the need for QEC is imminent. Most errors in quantum systems depend on the system's specific physical nature. Below are some of the generally occurring errors in quantum systems. 

\begin{enumerate}[label =(\roman*)]
    \item \textit{\textbf{Coherent quantum errors}:} In simpler words, these errors refer to errors that occur because of incorrectly applied gates. This would mean that if a system is assumed to be governed by a Hamiltonian $H$ when in reality it is governed by a Hamiltonian $H'$ \cite{devitt2013quantum}. Such a discrepancy in the system behavior is classified as coherent errors or systematic control errors. These errors are intrinsic to the qubit and thus can be mitigated using coherent techniques like composite pulse sequences (as defined in the following:  \cite{brown2004arbitrarily, jones2009composite}).

    We can understand this in more detail by looking at the example from  \cite{devitt2013quantum}. We assume that the incorrect characterization in the controls leads to a gate to not be $\sigma_I$, but rather introduces a small angle of rotation about $X-$axis of the Bloch sphere. Thus,
    \begin{equation}
        \small
        |\psi\rangle_{\text{final}} = \prod^N e^{i\epsilon\sigma_x}|0\rangle = \cos(N\epsilon)|0\rangle+i\sin(N\epsilon)|1\rangle.
        \label{eq:rotation}
    \end{equation}
    Now, if we measure the system in $|0\rangle$ and $|1\rangle$ basis, the probability of measuring the system in these would be, 
    \begin{align}
    P(|0\rangle) &= \cos^2(N\epsilon) \approx 1 - (N\epsilon)^2, \nonumber\\
    P(|1\rangle) &= \sin^2(N\epsilon) \approx (N\epsilon)^2.
    \end{align}
    Thus, we see a small probability of error, $p_\text{error} = (N\epsilon)^2$. It's small as $N\epsilon\ll 1$. 
    \vspace{0.2cm}
    \item \textit{\textbf{Decoherence Error}:} This is another important error that the qubit system happens. We can understand this error by using another simple example from  \cite{devitt2013quantum}. Let's consider an environment with two basis states, $|e_0\rangle$ and $|e_1\rangle$, which follows the following completeness relations, 
    \begin{align}
    \langle e_i | e_j \rangle &= \delta_{ij}, \\
    \ket{e_0}\bra{e_0} + \ket{e_1}\bra{e_1} &= I.
    \label{eq:decoherence}
    \end{align}
    Now, we assume that the environment is initialized in the state, $\ket{E}=\ket{e_0}$, and coupled to the system, 
    \begin{equation}
    \small
        HIH\ket{0}\ket{E}=\frac{1}{2}(\ket{0}+\ket{1})\ket{e_0}+\frac{1}{2}(\ket{0}-\ket{1})\ket{e_1}.
    \end{equation}
    We can write the density matrix, $\rho_f=(HIH\ket{0}\ket{E})(\bra{E}\bra{0}HIH)$ \cite{devitt2013quantum}. Without any knowledge of environmental degrees of freedom, the trace of the matrix is given as, 
    \begin{equation}
    \text{Tr}_E(\rho_f) = \frac{1}{2}(\ket{0}\bra{0}+\ket{1}\bra{1}).
    \end{equation}
    which gives us a mixed state. Measurement of this system will return $\ket{0}$ and $\ket{1}$ with $50\%$ probability each. The coupling to the environment removed all the coherences between the $\ket{0}$ and $\ket{1}$ states and the second Hadamard transform, in- tended to rotate $(\ket{0}+\ket{1})/\sqrt{2} \to \ket{0}$, has no eﬀect on the qubit state \cite{devitt2013quantum}.

    \item \textit{\textbf{Photon Loss Error}:} This is one of the most significant challenges in the photon-based quantum communication. In  \cite{czerwinski2022statistical}, the authors present a statistical analysis of photon loss in fiber-optic-based communications. Photon loss is modeled by the \textit{Beer–Lambert law}, describing the exponential decay of light intensity \( I(L) \) over distance \( L \):
    \begin{equation}
        I(L) = I_0 e^{-\lambda L},
    \end{equation}
    where:
    \begin{itemize}
        \item \( I(L) \): Intensity of transmitted light after distance \( L \),
        \item \( I_0 \): Initial light intensity,
        \item \( \lambda \): Rate parameter, related to the attenuation coefficient \( \alpha \).
    \end{itemize}
    
    The probability of photon loss in a quantum communication system is derived as:
    \begin{equation}
        P_{\text{loss}} = 1 - e^{-\lambda L},
    \end{equation}
    where \( e^{-\lambda L} \) represents the survival probability of a photon traveling distance \( L \). The loss can be modeled on a qubit state as follows, 
    \begin{equation}
    |1\rangle \to \sqrt{1-\eta}|1\rangle + \sqrt{\eta}|0\rangle,
    \end{equation}
    where,
    \begin{itemize}
        \item \( \eta \) represents the photon loss probability,
        \item \( \sqrt{1-\eta} \) is the survival amplitude,
        \item \( \sqrt{\eta} \) is the loss amplitude leading to the \( |0\rangle \) state.
    \end{itemize}

    \item \textit{\textbf{Noisy Intermediate-Scale Quantum (NISQ) Errors}:} Noisy Intermediate-Scale Quantum (NISQ) devices, characterized by having tens to hundreds of qubits, operate without full fault tolerance. These devices are susceptible to various error sources that collectively degrade the fidelity of quantum computations \cite{preskill2018quantum}.

\begin{itemize}

    \item \textit{Gate Errors}: Quantum gates are implemented with a finite fidelity due to decoherence and environmental noise. For example, a single-qubit rotation \( R_x(\theta) = e^{-i\theta\sigma_x/2} \) may have a small angle error:
    \begin{equation}
        \theta' = \theta + \delta\theta,
    \end{equation}
    where \( \delta\theta \) introduces coherent or systematic errors that accumulate over time.
    
    \item \textit{Finite Coherence Times}: Qubits have limited lifetimes, represented by \( T_1 \) (energy relaxation time) and \( T_2 \) (dephasing time). These limitations affect gate fidelity, as operations must complete within these time scales.
\begin{itemize}
    \item         $T_1$: \text{Decay from } $|1\rangle \to |0\rangle$, 
    \item $T_2$: \text{Loss of phase coherence between } $|0\rangle \text{ and } |1\rangle$.
\end{itemize}
    For example, after a time \( t > T_2 \), the density matrix of a qubit becomes:
    \begin{equation}
        \rho = \frac{1}{2}(|0\rangle\langle 0| + |1\rangle\langle 1|),
    \end{equation}
    indicating a complete loss of coherence.
    
    \item \textit{Drift in Calibration}: Over time, hardware parameters (e.g., control pulse amplitudes and frequencies) drift, leading to systematic errors. This drift is modeled as:
    \begin{equation}
        H_{\text{actual}} = H_{\text{ideal}} + \Delta H,
    \end{equation}
    where \( \Delta H \) represents the deviation due to miscalibrated controls.
    
    \item \textit{Systematic Bias in Readout Processes}: Measurement errors occur due to imperfections in the readout process. For instance, a qubit in state \( |0\rangle \) may be misread as \( |1\rangle \) with probability \( p_{\text{error}} \):
    \begin{equation}
        P(\text{readout error}) = p_{\text{error}}.
    \end{equation}
    Mitigating these errors requires readout error correction techniques, such as mapping \( |0\rangle \to |0\rangle \) and \( |1\rangle \to |1\rangle \) with higher fidelity.
    
    \end{itemize}
   
\end{enumerate}
These are some of the major errors in quantum communication. We would now go over some of the major works in quantum error correction. We would go over the breakthrough works such as the Calderbank-Shor-Steane (CSS) code, and recent studies with several quantum communication protocols.

Before we go into the details of QECCs, it's important to point out that most of these codes use certain techniques from classical coding theory. However, one of the difficulties while translating some of the classical methods into a quantum version is the no-cloning theorem for quantum states, as it prohibits us from defining a unitary operator, $U_c$, which carries out the following, 
\begin{equation}
    U_c(\ket{\psi}\otimes \ket{0}) \to \ket{\psi}\otimes \ket{\psi},
    \label{eq:clone}
\end{equation}
where, $\ket{\psi}$ is the quantum state to be cloned. This is a problem because most of classical coding theory assumes that data can be arbitrarily duplicated \cite{roffe2019quantum}. A quantum error correction code is represented in the following way,
\begin{equation}
    Q:[[n,k,d]],
\end{equation}
where,
\begin{itemize}
    \item $Q$: the quantum error correction code
    \item $n$: the number of the physical qubits used to store the information.
    \item $k$: the number of logical qubits protected by this code.
    \item $d$ is the distance of the code, i.e., if $d=3$ then the code protects against 1 flip or phase flip error occurring.
\end{itemize}
\subsection{Three Qubit Error Correction Code}
The three-qubit repetition code encodes one logical qubit into three physical qubits, providing protection against a single bit-flip error but not phase flips \cite{devitt2013quantum}.  The logical states are  
\begin{equation}
  \ket{0}_L = \ket{000}, 
  \qquad 
  \ket{1}_L = \ket{111},
\end{equation}
so that an arbitrary qubit $\ket{\psi}=\alpha\ket{0}+\beta\ket{1}$ is mapped to 
\begin{equation}
  \ket{\psi}_L = \alpha\ket{000} + \beta\ket{111}.
\end{equation}
The code distance $d=3$ implies it can correct up to $t=\lfloor(d-1)/2\rfloor=1$ bit‐flip error.  Error syndromes are extracted via two ancilla qubits, whose measurement uniquely identifies which physical qubit, if any, was flipped (Table~\ref{tab:error_states}).  Modeling coherent gate errors as $U=\cos\epsilon\,\sigma_I + i\sin\epsilon\,\sigma_x$, one finds the no‐error fidelity  
\begin{align}
  F_{\text{no\,error}} 
  &= \frac{\cos^6\epsilon}{\cos^6\epsilon + \sin^6\epsilon} 
  \approx 1 - \epsilon^6,
\end{align}
occurring with probability $1-3\epsilon^2 + O(\epsilon^4)$, and the detected‐error fidelity  
\begin{align}
  F_{\text{error}} 
  &= \frac{\cos^4\epsilon\,\sin^2\epsilon}{\cos^4\epsilon\,\sin^2\epsilon + \sin^4\epsilon\,\cos^2\epsilon} 
  \approx 1 - \epsilon^2,
\end{align}
with probability $3\epsilon^2 + O(\epsilon^4)$ \cite{devitt2013quantum}.  Since $\epsilon\ll1$, the encoded qubit maintains higher fidelity than an non-encoded one.  

\begin{table}[!htpb]
    \centering
    \renewcommand{\arraystretch}{1.5} 
    \begin{tabular}{|c|l|}
        \hline
        \textbf{Error Location} & \textbf{Final State} \\ \hline
        No Error & \(\alpha |000\rangle |00\rangle + \beta |111\rangle |00\rangle\) \\ \hline
        Qubit 1  & \(\alpha |100\rangle |11\rangle + \beta |011\rangle |11\rangle\) \\ \hline
        Qubit 2  & \(\alpha |010\rangle |10\rangle + \beta |101\rangle |10\rangle\) \\ \hline
        Qubit 3  & \(\alpha |001\rangle |01\rangle + \beta |110\rangle |01\rangle\) \\ \hline
    \end{tabular}
    \caption{Final state of a five-qubit system ($3$ from the encoding, and $2$ ancilla qubits before syndrome measurement reflects either no error or a single $X$-error on one of the qubits. The last two qubits represent the ancilla's state. Each possible error produces a unique measurement result (syndrome) from the ancilla qubits, enabling the identification of the error location. Based on the syndrome, a classically controlled ($\sigma_x$) gate is applied to the data block, ensuring error correction. This process leverages the unique syndrome for each error to maintain the integrity of the data qubits. Based on the work by Devitt et al \cite{devitt2013quantum}.}
    \label{tab:error_states}
\end{table}

\subsection{Stabilizer Formulation}
The stabilizer formalism \cite{gottesman1996class,gottesman1997stabilizer} encodes $k$ logical qubits into $n$ physical qubits by specifying an abelian subgroup $S$ of the $n$-qubit Pauli group $\mathcal{P}_n$ that does not contain $-I$.  Every $M\in S$ satisfies
\begin{equation}
  [M_i,M_j]=0,\quad \forall\,M_i,M_j\in S,
  \qquad -I\notin S.
\end{equation}
The code space 
\begin{equation}
  \mathcal{C}=\{\ket\psi:M\ket\psi=+\ket\psi,\;\forall M\in S\}
\end{equation}
has dimension $2^k$, where $k=n-r$ and $r$ is the number of independent stabilizer generators.  This framework generalizes classical linear codes: notable examples are the 5-qubit “perfect” code \cite{laflamme1996perfect}, the 7-qubit Steane code \cite{steane1996error}, and topological surface codes \cite{bravyi1998quantum}, each offering fault-tolerant protection against single-qubit errors.

\subsection{$5$-Qubit Error Correction Code}
The 5-qubit quantum error correction code, also known as the \textit{perfect code}, is the smallest quantum code capable of correcting an arbitrary single-qubit error. It encodes a single logical qubit into five physical qubits and protects against errors, including bit-flip, phase-flip, and combinations of both \cite{laflamme1996perfect, shor1995scheme}. This code is constructed using the stabilizer formalism, where logical operations are defined to preserve the encoded states.

The logical states \( |0_L\rangle \) and \( |1_L\rangle \) are encoded as superpositions of five physical qubits, satisfying specific stabilizer constraints:
\begin{align}
\centering
    \small
    \langle S_i \rangle &= +1 \quad \forall i, \\
    S_i &\in \{XZZXI, IXZZX, XIXZZ, ZXIXZ\}.
\end{align}

Here, \( X \), \( Z \), and \( I \) represent the Pauli operators and identity, respectively. Any single-qubit error maps the encoded state to an orthogonal subspace, allowing the syndrome measurement to detect and correct the error,
\begin{equation}
    S_i |\psi\rangle = (-1) |\psi\rangle \quad \text{(error detected)}.
\end{equation}
The 5-qubit code requires minimal overhead (only 5 physical qubits) and can correct any single-qubit error. However, it is sensitive to decoherence due to high connectivity requirements.

\subsection{$7-$Qubit Error Correction Code}
The 7-qubit Steane code \cite{steane1996multiple,steane1996error} is a CSS code built from the classical $[7,4,3]$ Hamming code. It encodes one logical qubit into seven physical qubits and corrects any single-qubit bit-flip or phase-flip error. The logical basis states are
\begin{align}
  \ket{0_L} &= \frac{1}{\sqrt{8}}\sum_{c\in C_1}\ket{c}, 
  &\ket{1_L} &= \frac{1}{\sqrt{8}}\sum_{c\in C_1}\ket{c\oplus1111111},
\end{align}
where $C_1$ is the set of Hamming codewords. Its stabilizer generators are
\begin{gather}
  S_1 = IIIXXXX,\quad S_2 = IXXIIXX,\quad S_3 = XIXIXIX,\\
  S_4 = IIIZZZZ,\quad S_5 = IZZIIZZ,\quad S_6 = ZIZIZIZ,
\end{gather}
and valid codewords satisfy $S_i\ket{\psi_L}=+\ket{\psi_L}$. Measuring the stabilizers reveals a unique error syndrome, which is corrected by applying the corresponding Pauli operator. With only seven physical qubits per logical qubit, the Steane code is foundational for fault-tolerant quantum computing, though its resource overhead limits scalability to larger systems.

\subsection{$9-$Qubit Error Correction Code}
The $9-$qubit error correction code was given by Peter Shor, and it was one of the first QECC that can correct one bit-flip and phase-flip error, thus can correct any arbitrary single-qubit error.  It encodes a single logical qubit in nine physical qubits and protects against bit-flip and phase-flip errors by combining repetition and entanglement-based encoding  \cite{shor1995scheme, 548464}. This code is a concatenated code where logical qubits are protected by nested layers of simpler error-correcting codes.

The encoding of logical qubits into physical qubits is the equipartition of all three qubit states. The logical states \( |0_L\rangle \) and \( |1_L\rangle \) are encoded as:
\begin{align}
    |0_L\rangle &= \frac{1}{\sqrt{8}} \left(|000\rangle + |111\rangle\right)^{\otimes 3}, \nonumber\\
    |1_L\rangle &= \frac{1}{\sqrt{8}} \left(|000\rangle - |111\rangle\right)^{\otimes 3}.
\end{align}

This encoding combines a 3-qubit repetition code for correcting bit-flip errors:
\begin{equation}
    |0\rangle \to |000\rangle, \quad |1\rangle \to |111\rangle,
\end{equation}
with a phase encoding scheme to correct phase-flip errors:
\begin{equation}
    |\pm\rangle = \frac{1}{\sqrt{2}} \left(|0\rangle \pm |1\rangle\right).
\end{equation}

The 9-qubit code ensures that any single-qubit error maps the encoded state to a distinct error syndrome, enabling error detection and correction. While the 9-qubit code demonstrates conceptual simplicity and foundational importance, it requires significant overhead with nine physical qubits per logical qubit, limiting scalability for larger systems.

\subsection{CSS Error Correction Code}
The Calderbank-Shor-Steane (CSS) Code is a class of quantum error correction codes based on linear coding theory.  We start with choosing two linear code basis, $C_1$ ($[n_1, k_1, d_1]$) and $C_2$ ($[n_2, k_2, d_2]$) with $C_2^\perp \subseteq C_1 \quad \text{and} \quad n \leq k_1 + k_2$. We can construct a quantum code using this,  \cite{calderbank1996good, jha2024joint}
\begin{equation}
    Q:[[n, k=k_1+k_2, d=\min(d_1,d_2)]]
\end{equation}
Now, we can write the basis states based on the above construction as follows,
\begin{equation}
    \small
    \left\{ |c_w\rangle = \frac{1}{2^{\frac{n-k_1}{2}}} \sum_{w \in C_1^\perp} |w + v\rangle, \, w \in C_2 / C_1^\perp \right\},
\end{equation}
The coding for the logical bases is done on the cosets of these two code basis. Logical qubits are encoded using the cosets of \( C_2 \) in \( C_1 \). The logical states \( |0_L\rangle \) and \( |1_L\rangle \) are expressed as:
\begin{equation}
   |0_L\rangle = \frac{1}{\sqrt{|C_2|}} \sum_{c \in C_2} |c\rangle, \quad
   |1_L\rangle = \frac{1}{\sqrt{|C_2|}} \sum_{c \in C_2} |c + d\rangle,
\end{equation}
where \( d \) is a vector in \( C_1 \setminus C_2 \), i.e., not in \( C_2 \).

The stabilizer formalism of the CSS code facilitates error detection and correction. Bit-flip errors are detected using the parity check matrix of \( C_1 \):
\begin{equation}
    S_X = \{ X^{v} : v \in C_1^\perp \},
\end{equation}
where \( C_1^\perp \) is the dual of \( C_1 \), and \( X^v \) applies the Pauli \( X \)-operator on the qubits indexed by \( v \). Phase-flip errors are detected using the parity check matrix of \( C_2^\perp \):
\begin{equation}
    S_Z = \{ Z^{w} : w \in C_2^\perp \}.
\end{equation}
Now that we've laid down the basic theoretical formulation of the major Quantum Error Correction Codes (QECCs), we will go over some of the recent experimental and simulation works on the actual application of these techniques to the Quantum Communication framework.

\subsection{Relevant Works on QECCs}
In the past decade and a half, the development of QKD networks has resulted in increasing interest in the development of QECCs. 

\begin{enumerate}[label =(\roman*)]
    \item Michael et al., present a new class of error correction in bosonic modes in the form of a \textit{Binomial Error Correction Code}. These codes mainly address the issues such as photon loss, photon gains, and dephasing errors in single bosonic modes \cite{michael2016new}. Unlike traditional qubit-based quantum codes, these codes use finite superpositions of Fock states weighted by binomial coefficients to encode logical qubits, ensuring compatibility with continuous-variable systems. This is done with regards to the lowest $2^M$ Fock states of a single mode of a resonator, under the assumption of a damped harmonic oscillator with photon loss \cite{devoret2013superconducting}.

    The authors generalize the class of error correction codes to protect against the error set,
    \begin{equation}
    \small
        \mathcal{E} = \{ \hat{I}, \hat{a}, \hat{a}^2, \ldots, \hat{a}^k, \hat{a}^\dagger, \hat{a}^{\dagger 2}, \ldots, (\hat{a}^\dagger)^q, \hat{n}, \hat{n}^2, \ldots, \hat{n}^D \},
        \label{error}
    \end{equation}
    which can corrected upto $L$ photon losses, $G$ photon gain errors, and $D$ dephasing errors. This new generalized class of error correction codes can correct for errors in eq(\ref{error}), 
    \begin{equation}
    |W_{\uparrow/\downarrow}\rangle = \frac{1}{\sqrt{2^N}} \sum_{p=0}^{N+1} \sqrt{\binom{N+1}{p}} |p(S+1)\rangle,
    \end{equation}
    where,
    \begin{itemize}
        \item \( N \) specifies the number of correctable photon-loss errors
        \item \( S \) determines the spacing of Fock states.
    \end{itemize}
     This encoding satisfies the Knill-Laflamme quantum error-correction criteria, enabling protection against photon loss, gain, and dephasing errors. Recovery is achieved through unitary operations and measurements on the bosonic mode. The recovery of the states are done using a series of unitary transformations.

     The main results show that the construction using binomial codes effectively mitigates the above-mentioned errors, such as photon loss, photon gain, and dephasing errors. The uncorrectable error rate for photon loss in a Binomial Error-Correcting Code is given by:
    
        \begin{equation}
                P_{L+1} \sim \kappa (\kappa \delta t)^L \left( \frac{\langle (\hat{a}^\dagger)^{L+1} \hat{a}^{L+1} \rangle}{(L+1)!}\right).
        \end{equation}
    where,
    \begin{itemize}
        \item \( P_{L+1} \): Probability of \( L+1 \) photon loss errors occurring.
        \item \( \kappa \): Photon loss rate, which characterizes the rate of dissipation in the bosonic mode.
        \item \( \delta t \): Time step during which the photon loss occurs.
        \item \( L \): Number of photon loss events under consideration.
        \item \( \hat{a} \): Annihilation operator, which removes a photon from the mode.
        \item \( \hat{a}^\dagger \): Creation operator, which adds a photon to the mode.
        \item \( \langle (\hat{a}^\dagger)^{L+1} \hat{a}^{L+1} \rangle \): Expectation value representing the system’s state after \( L+1 \) photon loss events.
        \item \( (L+1)! \): accounts for the combinatorial contributions of the error events.
    \end{itemize}
    This study provides a comprehensive theoretical framework of error correction codes for bosonic states. However, there are a few limitations of this work. 
    \begin{itemize}
        \item The effectiveness of binomial codes diminishes for high-order errors $(L+1$ losses), as the uncorrectable error rate still grows with $\kappa^L$.
        \item The need for highly accurate unitary transformation and measurements on bosonic modes might be physically challenging in itself. This might serve as a practical limitation for this work. 
    \end{itemize}

    \item Reed et al., present experimental scenarios to realize the three-qubit quantum error correction code, to detect and correct single-qubit errors. This implementation demonstrates bit-flip and phase-flip error correction using a superconducting circuit architecture, which leverages the coherence and control of transmon qubits. This work is a foundational step toward scalable quantum error correction \cite{reed2012realization}. 

    The Travis-Cummings Hamiltonian describing the system in the study with four transmon qubits is described as, 
    \begin{align}
    H &= \hbar \omega_c a^\dagger a 
    + \hbar \sum_{q=1}^4 \Bigg( \sum_{j=0}^N \omega_{0j}^{(q)} |j\rangle_q \langle j|_q \nonumber \\
    &\quad + (a + a^\dagger) \sum_{j,k=0}^N g_{jk}^{(q)} |j\rangle_q \langle k|_q \Bigg).
    \end{align}
    where, 
    \begin{itemize}
    \item \( \hbar \): Planck's reduced constant.
    \item \( \omega_c \): Bare cavity frequency, describing the resonant frequency of the cavity mode.
    \item \( a, a^\dagger \): Annihilation and creation operators for the cavity mode.
    \item \( \omega_{0j}^{(q)} \): Transition frequency for transmon \( q \) from the ground state to the \( j \)-th excited state.
    \item \( g_{jk}^{(q)} = g_q n_{jk} \): Coupling matrix element between transmon \( q \) and the cavity mode, where:
    \begin{itemize}
        \item \( g_q \): Bare qubit-cavity coupling strength.
        \item \( n_{jk} \): Coupling matrix element for transitions between states \( |j\rangle \) and \( |k\rangle \).
    \end{itemize}
    \item \( |j\rangle_q, |k\rangle_q \): Basis states of transmon \( q \) for the \( j \)-th and \( k \)-th energy levels.
    \item \( E_{Cq} \): Charging energy of the transmon qubit.
    \item \( E_{Jq} \): Josephson energy of the transmon qubit, which depends on the flux threading the SQUID loop.
    \item \( E_{Jq} = E_{Jq}^{\text{max}} |\cos(\pi \Phi_q / \Phi_0)| \): Flux-dependent Josephson energy \cite{koch2007charge}.
    \end{itemize}

    The experimental setup is described as below,
    \begin{itemize}
    \item \textit{Encoding}: A single logical qubit is encoded into a three-qubit entangled state, such as the Greenberger–Horne–Zeilinger (GHZ) state.
    \item \textit{Error Syndrome Measurement}: Errors are mapped to ancillary qubits through reversible encoding processes, which are then corrected without disturbing the encoded quantum state.
    \item \textit{Toffoli Gate Implementation}: A high-fidelity conditional-conditional NOT (CCNot or Toffoli) gate, essential for error correction, is implemented using the third excitation level of a transmon qubit. The gate operation achieved a fidelity of 85\%.
    \item \textit{Error Correction}: Errors are corrected by decoding the error syndrome back to the ancillas and applying the Toffoli gate, ensuring first-order insensitivity to errors.
    \end{itemize}

    The main result of this study was presented as process fidelity values. The process fidelity is modelled as, 
    \begin{equation}
        F=1-3p^2+2p^3,
    \end{equation}
    where $p$ is the single qubit error rate. The process fidelity for the bit-flip and phase-flip correction schemes was 76\%, demonstrating the suppression of linear error terms (first-order insensitivity). The implementation of the Toffoli gate achieved 85\% fidelity for classical action and 78\% fidelity for quantum process action. The QEC protocol reduced the error rate quadratic to the single-qubit error rate, showcasing the effectiveness of the correction mechanism. There are few limitations of this work, 
    \begin{itemize}
        \item The implementation is constrained by the coherence times of superconducting qubits, with lifetimes around $T_1=0.7-1.3\mu s$.
        \item The gate fidelity, while significant, might not be sufficient for large-scale fault-tolerant quantum computing.
        \item Only two error models are implemented; more complex error models such as amplitude damping, dephasing errors, etc. 
    \end{itemize}
    This work serves as a quite detailed work implementing the simplest form of QECC, and can be important for future works on implementing more complex QECCs on superconducting qubits. 

    \item Cai et al., present an experimental realization of repeated quantum error correction cycles using the distance-three surface code. The surface code, a stabilizer code based on topological features of a 2D qubit lattice, is highly tolerant to noise and crucial for fault-tolerant quantum computing \cite{cai2021bosonic}. By encoding one logical qubit into nine data qubits and using eight auxiliary qubits for stabilizer measurements, the authors implement a scalable QEC cycle capable of detecting and correcting both bit- and phase-flip errors. The authors implement the surface code in a superconducting circuit, leveraging advanced techniques to,
    \begin{itemize}
        \item Measure stabilizer syndromes using controlled-phase (CZ) gates and simultaneous parallel execution of stabilizers.
        \item Achieve a fast QEC cycle time of $1.1 \mu s$.
        \item Detect and mitigate errors, including leakage errors, using high-fidelity readout and minimum-weight perfect matching for syndrome decoding.
    \end{itemize}
    The study demonstrates logical state preservation through repeated QEC cycles and analyzes the logical fidelity and error rates. The study setup can be described as follows,
    \begin{itemize}
        \item \textbf{Logical State Encoding:}  The logical qubit is encoded using nine physical data qubits. The logical basis states are defined as:
    \begin{align}
        |0_L\rangle &= \frac{1}{\sqrt{2}} \Big(|000\rangle + |111\rangle\Big), \nonumber \\
        |1_L\rangle &= \frac{1}{\sqrt{2}} \Big(|000\rangle - |111\rangle\Big).
    \end{align}
    These states span the logical subspace protected by the surface code.
    \item \textbf{Stabilizer Formalism:}  
    The stabilizer generators for the surface code are defined to constrain the logical state and detect errors. For a distance-three surface code, the stabilizers include:
\begin{align}
    S_X^{(1)} &= X_1 X_2, \nonumber\\
    S_X^{(2)} &= X_2 X_3, \nonumber\\
    S_X^{(3)} &= X_3 X_4,
\end{align}

\begin{align}
    S_Z^{(1)} &= Z_1 Z_2, \nonumber\\
    S_Z^{(2)} &= Z_2 Z_3, \nonumber\\
    S_Z^{(3)} &= Z_3 Z_4.
\end{align}

    Each stabilizer acts on a subset of qubits to measure syndromes that indicate the presence of errors.
    \item \textbf{Syndrome Measurement:}  
    Stabilizer measurements project the state into eigenspaces of the operators \( S_X \) and \( S_Z \), with outcomes \( \pm 1 \). An error \( E \) is detected if:
    \begin{align}
        S_X |E\rangle &= -1 |E\rangle \quad \text{or} \quad S_Z |E\rangle = -1 |E\rangle.
    \end{align}
    \item \textbf{Error Detection and Correction:}  
    Errors are interpreted based on the syndrome measurements:
    \begin{itemize}
        \item For a detected error, a \textit{minimum-weight perfect matching algorithm} is used to determine the most probable error configuration.
        \item A correction operator \( E_c \) is applied to restore the logical state:
        \begin{align}
            |L\rangle \to E_c |L\rangle.
        \end{align}
    \end{itemize}
    \item \textbf{Gate Operations:}  Stabilizer measurements are implemented using controlled-phase (CZ) gates and single-qubit rotations. The ancillary qubits interact with the data qubits to extract syndromes.
    
    \textbf{Logical Error Rate:}  
    The logical error rate \( \epsilon_L \) is determined by the physical error rate \( p \) and the distance \( d \) of the code:
    \begin{align}
        \epsilon_L \sim p^{(d+1)/2}.
    \end{align}
    For the distance-three surface code, \( \epsilon_L\propto p^2 \), highlighting its second-order fault-tolerance.

    \end{itemize}
    The logical qubit states \( |0_L\rangle, |1_L\rangle, |+\rangle_L, |-\rangle_L \) are initialized with fidelity exceeding $99\%$. The logical coherence time \( T_2 \) and relaxation time \( T_1 \) are approximately $18.2\ \mu s$ and $16.4\ \mu s$, respectively. The logical error probability per cycle (\( \epsilon_L \)) is around $3\%$, matching simulated values. The primary limitation of the error rate is due to physical qubit coherence times and gate fidelities. Scalability is demonstrated through a pipelined approach to stabilizer measurement, enabling potential scalability to larger surface code distances. This work was an extension to the work presented in  \cite{michael2016new}, and it serves as a good baseline for future works concerning different error correction frameworks for bosonic modes. 
    
    \item Sivak et al. demonstrate a breakthrough in quantum error correction (QEC) by stabilizing a logical qubit whose coherence surpasses that of the underlying physical qubits. The results of the study indicate that QEC prolongs the quantum information lifetime beyond the "break-even" point where the logical qubit outperforms any single physical qubit in the system \cite{sivak2023real}. The following are the major proposals by the authors, 
    \begin{itemize}
        \item Implementation of the Gottesman-Kitaev-Preskill (GKP)  \cite{gottesman2001encoding} bosonic encoding in superconducting circuits.
        \item Use of a novel ``trickle-down" dissipation scheme for error correction, which simplifies control overhead while still correcting most probable errors efficiently. (This was proposed originally in  \cite{de2022error, ryan2021realization, abobeih2022fault, xue2022quantum})
        \item Integration of reinforcement learning (RL) to optimize QEC parameters in real time, adapting to the error channels and imperfections in the system.
    \end{itemize}
    The logical qubit stabilization achieved significant coherence times and error correction efficiency as listed below,

    \begin{itemize}
        \item Logical coherence times were measured as:
        \begin{align}
            T_X = T_Z &= 2.20 \, \text{ms}, \nonumber\\
            T_Y &= 1.36 \, \text{ms}.
        \end{align}
    
        \item Logical error rates were calculated as:
        \begin{align}
            p_X = p_Z &= 1.81 \times 10^{-3}, \nonumber\\
            p_Y &= 4.3 \times 10^{-4}.
        \end{align}
    
        \item A coherence gain of:
        \begin{align}
            G = 2.27
        \end{align}
        was achieved, surpassing the "break-even" point, marking a milestone in practical quantum error correction.
    
        \item The reinforcement learning-based optimization improved quantum error correction parameters, leading to 97\% error correction success for most probable errors.
    \end{itemize}

    This work offers a new approach to error correction techniques. However, there are a few limitations as pointed out below,
    \begin{itemize}
        \item The system's logical performance was occasionally degraded by interactions with spurious degrees of freedom, such as higher transmon states or environmental defects.
        \item The ``trickle-down" dissipation scheme, while reducing overhead, is less efficient for correcting rare, large errors.
        \item Ancilla qubit errors, particularly bit flips, were identified as a significant source of logical decoherence.
    \end{itemize}

    \item Postler et al., introduce the implementation of multiple rounds of fault-tolerant (FT) quantum error correction (QEC) based on the Steane method in a trapped-ion quantum processor. The Steane QEC utilizes a transversal controlled-NOT (CNOT) gate to minimize interaction between logical qubits and auxiliary qubits, reducing error propagation \cite{postler2024demonstration}. The method is tested on the 1D repetition codes and the $7$-qubit color code, showcasing fault-tolerant quantum computation. The transversal CNOT operation copies errors onto the auxiliary logical qubit,

    \begin{align}
    C_{\text{NOT}} \left( X_i^{\text{data}} \otimes I \right) 
    |\psi\rangle_L \otimes |+\rangle_L 
    &= \nonumber\\ X_i^{\text{data}} |\psi\rangle_L \otimes X_i^{\text{aux}} |+\rangle_L.
    \end{align}

    Logical fidelities are evaluated as probabilities of recovering the encoded state within the correction capabilities of the code.
    This work is one of the initial works exploring the experimental implementation of experimental application of Steane syndrome extraction for error correction. This method should reduce errors introduced by the coupling mechanism and simplify syndrome extraction by using projective measurements.

    The implementation demonstrated improvements in logical fidelities for bit-flip and phase-flip repetition codes of distances 3 and 5, as well as for the $7$-qubit color code. The Steane QEC outperformed flag-based QEC approaches under similar conditions, particularly for logical input states $\ket{0}_L$ The advantage of Steane QEC becomes more pronounced in the regime dominated by two-qubit gate errors, as demonstrated through Monte Carlo simulations. Some of the limitations of this work can be listed as follows, 
    \begin{itemize}
        \item Mid-circuit measurements introduce dephasing and depolarizing noise on data qubits, impacting overall fidelity.
        \item Increased gate overhead for codes with higher distances, requiring careful error modeling to maintain fault-tolerance.
        \item Experimental setup constraints, such as limited qubit connectivity and imperfect hardware, restrict scalability.
    \end{itemize}
    \end{enumerate}

\noindent To sum up, these theoretical constructions and experimental demonstrations illustrate the rapid progress in quantum error correction—from bosonic binomial codes and small‐scale superconducting realizations to more complex topological surface codes, GKP bosonic encoding, and fault‐tolerant trapped‐ion implementations.  However, there are several challenges.  First, most codes have been validated either in isolation or under highly idealized noise models; integrating them into end‐to‐end QKD networks will require tailoring code parameters to the specific loss, dephasing, and side‐channel error profiles encountered in real network links.  Second, scalable, low‐overhead recovery operations (e.g., syndrome extraction, and adaptive control) must be co‐designed with photonic or cryogenic platforms to meet stringent timing and fidelity requirements.  Third, hybrid approaches, i.e., combining bosonic and qubit codes, or uniting QEC with classical post‐processing, offer promising paths to reduce overhead, but their performance under coherent and adversarial attacks is largely unexplored.

\section{Relevance of this Review}
\label{Sec:Relevance}
In recent times, amid the NISQ era, quantum augmented networks (QuANets) have been proposed as one of the feasible approaches to global-scale quantum network \cite{jha2024ml, jha2025towards}. These networks explore a hybrid approach of integrating quantum components strategically in classical networks to have an overall higher security than classical networks and higher efficiency than quantum networks. In a QuANet, classical and quantum payloads coexist within a single infrastructure, with quantum resources (e.g., qubit channels, single-photon detectors, quantum memories) reserved exclusively for sensitive or ``private” communications. As such, every element of the hybrid stack (ranging from application-layer privacy classification to physical-layer quantum transmission) must be informed by an intimate understanding of the vulnerabilities and countermeasures summarized in this review. The following are important points to be noted,

\begin{enumerate}
    \item Most of the QuANet proposals rely on quantum key distribution (QKD) or quantum-secure direct communication to protect high-value payloads. Yet without a careful appraisal of active and passive eavesdropping techniques—most notably Trojan-horse injections (Section \ref{Sec:trojan}) and PNS attacks (Section \ref{PNS})—the promise of information-theoretic security can be undermined in practice. For example, a Trojan-horse injection that exploits unmonitored back-scatter may leak phase‐reference information to Eve unless optical isolators, narrow-band filters, and real-time reflectometry are incorporated into transmitter and receiver modules. Likewise, decoy-state countermeasures—while effective against idealized PNS attacks—must be tuned to detect joint‐PNS variants that siphon photons bidirectionally. Furthermore, decoy-state countermeasures are not very practical solutions as these decrease the key rate significantly \cite{jha2024multi}. In a QuANet context, these insights will help directly in the hardware design of several key components such as quantum gateways and switches, ensuring that any node arbitrating between classical and quantum paths can also monitor for side‐channel leakage.
    \item QuANets are inherently subject to denial-of-service (DoS) attacks via jamming (Section \ref{Sec:jamming}), both quantum jamming attacks, and classical jamming attacks. Whether an adversary attempts to rotate polarization through a Faraday jamming field or injects broadband noise into free-space links, the net effect is an elevated quantum bit-error rate (QBER) or a suppressed Bell-inequality violation parameter ($S$), forcing legitimate users to abort key‐exchange sessions. By quantifying how an injected noise fraction $p$ suppresses fidelity ($F = 1 - p/2$) and drives 
    \begin{equation}
        \mathrm{QBER}_\text{induced} \;=\; 1 \;-\; \cos^2(\beta),
        \label{eq:qber1}
    \end{equation}
    One can derive threshold conditions under which quantum payloads are deemed insecure. In a QuANet, real-time QBER monitoring and adaptive routing logic (e.g., diverting affected qubits to re-transmission or fallback to classical encryption) become critical. In particular, the use of single-photon avalanche diode (SPAD) arrays or MDI-QKD architectures can help mitigate jamming by decoupling Eve’s injected noise from the legitimate quantum signal.

    \item The successful operation of a QuANet depends on robust error correction: quantum hardware have several losses, dephasing, or photon-gain errors over realistic fiber channels. The binomial and bosonic codes, the three-qubit and seven-qubit stabilizer codes, and topological codes such as the distance-three surface code all illustrate ways to recover from single- and multi-photon losses, as well as phase flips, with high probability. In a QuANet, where the quantum channel may share a single fiber with classical data (via wavelength-division multiplexing), crosstalk and channel uncertainty can increase error rates beyond those seen in dedicated QKD links. Hence, choosing an error-correction scheme with a sufficiently large distance (e.g., surface-code distance $d \ge 3$) and integrating fault-tolerant syndrome extraction will be mandatory to reach “break-even” logical-qubit coherence. Moreover, recent demonstrations of dissipation and machine-learning-aided parameter tuning can ensure that QuANet nodes adaptively optimize their error-correcting circuitry to several ambient noise.

    \item  From a network-architecture perspective, the review’s survey of security proofs under arbitrarily varying channels and the characterization of symmetrizability (i.e., whether a jammer can force $I(X;B\mid s) = I(X;B\mid s')$ for two different states $s,s'$) directly inform QuANet routing and coding policies. For instance, if a given fiber segment is known to be symmetrizable (zero deterministic secrecy capacity), the QuANet’s control plane must either reallocate quantum-secured flows to an alternate path or drop back to classical encryption for that segment. In practice, a security review helps QuANet architects identify which links can safely carry quantum payloads, how to schedule decoy-state intensities, and when to relegate traffic to classical fallback. 
\end{enumerate}

As QuANets consists of both classical and quantum components, a security analysis would consist of a combination of two proofs: (1) quantum security proofs, and (2) classical security proofs. Furthermore, to make QuANets a practically viable solution, we need to perform reliable error corrections in both the quantum and classical nodes. Together, these considerations underscore why a comprehensive security review is a prerequisite for any practical QuANet deployment. By aligning hardware-layer countermeasures, error-correcting codes, and adaptive control strategies, QuANets can achieve their promise of offering near-classical performance for non-sensitive traffic, while simultaneously elevating security for private communications to the information-theoretic level.

\section{Conclusion}
In this study, we provide a comprehensive review of the security aspect of Quantum Key Distribution. In the past two decades, QKD has been of utmost interest to researchers across the world. The interest is mainly due to the aspect of unconditional security offered by the QKD protocols due to the laws of quantum physics. However, there are other ways in which an eavesdropper can exploit the security offered by QKD. Thus, it becomes vital that we study the different kinds of attacks on quantum communication systems. In this review, we go over some of the legacy and modern QKD protocols and several theoretical security proofs for different QKD protocols.  Furthermore, we present an in-depth review of the existing works on three main classes of quantum attacks- photon-number splitting attacks (PNS), Trojan-horse attacks, and jamming attacks. Once we have discussed the security proofs and security threats to currently existing protocols, we provide an in-depth review of several quantum error correction codes, which become central to the discussion about the practicality of these QKD protocols. 

\section*{Disclosure statement}
The authors report there are no competing interests to declare.

\section*{Data availability statement}
All data generated or analyzed during this study are included in this published article [and its supplementary information files].

\section*{Funding Declaration}

\printbibliography

\section*{Author's Biography}
\bigskip
\noindent\textbf{Nitin Jha} received his BSc (Hons) in Physics from Ashoka University in 2023. He is currently pursuing his PhD at Kennesaw State University. His research interests lie in quantum networks, secure quantum communication, and network security. He’s exploring AI-based methods to enhance quantum-classical hybrid (quantum-augmented) networks. Nitin’s work has garnered multiple accolades at KSU and appears in top venues.

\bigskip
\noindent\textbf{Abhishek Parakh} is a Professor of Computer Science and Director of the Computer Science PhD Program at Kennesaw State University. He earned his PhD in Computer Science from Oklahoma State University in 2011. He previously served as Director of NebraskaCYBER and as the Mutual of Omaha Distinguished Chair of Information Science and Technology. His research spans quantum cryptography, cybersecurity, and trustworthy computing, with funding from NSF, DOD, NASA, NSA, and DOS.

\bigskip
\noindent\textbf{Mahadevan Subramaniam} is the Charles W. and Margre H. Durham Distinguished Professor and Chair of the Computer Science Department at the University of Nebraska Omaha. He earned his BS in Computer Science from BITS Pilani in 1986 and his MS and PhD from SUNY Albany in 1997. His work centers on formal methods for software engineering—model checking, SMT solvers, theorem provers—with applications in software evolution and automated repair.

\end{document}